\documentstyle[12pt]{article}
    \def\l{\lambda}
    \def\m{\mu}
    \def\a{\alpha}
    
    \def\o{\omega}
    \def\mod#1{|#1|}
\begin{document}

\begin{center}
{\Large \bf  Ground--$\gamma$ band coupling in heavy deformed nuclei
and SU(3) contraction limit}
\end{center}
\medskip

\begin{center}
{\large N. Minkov$^*$\footnote[1]{e-mail: nminkov@inrne.bas.bg},
S. B. Drenska$^*$\footnote[2]{e-mail: sdren@inrne.bas.bg},
P. P. Raychev$^{*}$\footnote[3]{e-mail: raychev@bgcict.acad.bg},
R. P. Roussev$^*$\footnote[4]{e-mail: rousev@inrne.bas.bg}
and Dennis Bonatsos$^\dagger$\footnote[5]{e-mail:
bonat@mail.demokritos.gr}\\
\medskip
$^*$  Institute for Nuclear Research and Nuclear Energy, \\
72 Tzarigrad Road, 1784 Sofia, Bulgaria\\
\medskip
$^\dagger$ Institute of Nuclear Physics, N.C.S.R. ``Demokritos'',\\
GR-15310 Aghia Paraskevi, Attiki, Greece}
\end{center}
\bigskip\bigskip

\begin{abstract}
We derive analytic expressions for the energies and $B(E2)$-transition
probabilities in the states of the ground and $\gamma$ bands of heavy
deformed nuclei within a collective Vector-Boson Model with  SU(3)
dynamical symmetry.  On this basis we examine the analytic behavior
of the SU(3) energy splitting and the B(E2) interband transition
ratios in the SU(3) contraction limits of the model.  The theoretical
analyses outline physically reasonable ways in which the
ground--$\gamma$ band coupling vanishes.  The experimental data on
the lowest collective states of even-even rare earth nuclei and
actinides strongly support the theoretical results.  They suggest
that a transition from the ground--$\gamma$ band coupling scheme to a
scheme in which the ground band is situated in a separate irreducible
representation of SU(3) should be realized towards the midshell
regions. We propose that generally the SU(3) group contraction
process should play an important role for such a kind of transitions
in any collective band coupling scheme in heavy deformed nuclei.

\end{abstract}
\bigskip\bigskip

PACS Numbers: 21.60.Fw, 21.60.Ev, 23.20.Js

\newpage
\section{Introduction}

An important advantage of the dynamical symmetry (DS) approach
\cite{Barut64,DGN65,Mukun65,Dash} in nuclear theory is the
possibility to describe consistently various collective bands of
heavy deformed nuclei \cite{Wea,p:sp6r,Af}. Generally, the DS concept
is based on the assumption that the physical system possesses a
``primary'' symmetry with respect to a given group, called DS group.
The Hamiltonian of the system reduces this symmetry to the group of
invariance of the system (which for the nuclear system coincides with
the angular momentum group SO(3)) and thus the energy spectrum is
generated \cite{Barut64}--\cite{Dash}. The Lie algebra of the DS
group is then reduced to the algebra of the group of invariance and
is referred to as spectrum generating algebra. The basic idea of DS
approach in heavy deformed nuclei is that their collective bands can
be united into one or several multiplets, appearing in this reduction
\cite{Wea,p:sp6r,Af}. It provides a natural way to study the
interaction between a particular couple of bands as well as the
attendant spectroscopic characteristics of nuclei.

Various classification schemes with band coupling have been developed
on the basis of DS approach. Well known models, such as the
Interacting Boson Model (IBM) \cite{IA}, the symplectic models
\cite{Rose1,Fill} and the Fermion Dynamical Symmetry Model
\cite{FDSM1}, provide a good overall description of nuclear
collective phenomena, covering the different regions of vibrational,
rotational and transitional nuclei.

On the other hand, some models, based on the SU(3) dynamical
symmetry, reproduce successfully the particular characteristics of
rotational bands in deformed nuclei.  Such models are the
Pseudo-SU(3) Model \cite{Jerry1}, which has microscopic
motivations, as well as the Vector-Boson Model (VBM) with SU(3)
dynamical symmetry \cite{p:descr,a:over,p:matr}, which allows a
relevant phenomenological treatment of the SU(3) multiplets in
nuclei.

While in the SU(3) limit of the IBM the possible irreducible
representations (irreps) ($\l, \m$) are restricted by the total
number of bosons describing the specific nucleus, in the VBM the
possible SU(3) irreps ($\l ,\m$) are not restricted by the underlying
theory.  However, it has been shown recently \cite{MDRRB97} that some
favored regions of ($\l ,\m$) multiplets in the VBM could be outlined
through the numerical analysis of the experimental data available for
the ground ($g$) and the $\gamma$- collective bands of even--even
deformed nuclei. (The favored multiplets provide the best model
descriptions.) As a result, a systematic behavior of the SU(3)
symmetry properties of rotational nuclei has been established in
terms of the VBM.  It suggests the presence of a transition between a
scheme, in which the $g$ and the $\gamma$ bands are coupled into one
and the same ($\l ,\m$) irrep and a scheme, where these two bands
belong to different irreps. In addition it has been supposed that the
fine systematic properties of rotational spectra could be interpreted
as a manifestation of a more general dynamical symmetry.

As a first step in the recovering of the dynamical mechanism causing
such a transition, one should study the way in which the SU(3)
symmetry is reduced in the ($\l ,\m$)- plane.  In particular, it is
of interest to reproduce the limits, in which the quantum numbers
$\l$ and $\m$ go to infinity, i.e.  the cases, in which the SU(3)
irreps are not finite anymore. These limits correspond to the so
called SU(3) contraction process, in which the algebra of SU(3) goes
to the algebra of the semi-direct product T$_{5}\wedge$SO(3), i.e.
$SU(3)\rightarrow T_{5}\wedge SO(3)$ (T$_5$ is the group of
5-dimensional translations generated by the components of
the SU(3)- quadrupole operators)
\cite{Gilmore,CBVR86,CDL88,RVC89,Mukerjee,JPD93}. Generally, the
contraction limit corresponds to a singular linear transformation of
the basis of a given Lie algebra. The transformed structure constants
approach a well-defined limits and a new Lie algebra, called
contracted algebra, results \cite{Gilmore} . The original and the
contracted algebra are not isomorphic.

On the above basis it is expected that in the SU(3) contraction limit
the space of the SU(3) irreps should undergo a respective limiting
transition. As a result the SU(3) multiplets should be disintegrated
to sets of various independent bands. It is, therefore, reasonable to
consider this limit as a natural way in which the band-mixing
interactions vanish. It is important to remark that the SU(3)
contraction process is a situation in which a compact group goes to a
non-compact one. Hence, one could try to interpret the vanishing
$g$--$\gamma$ band-mixing interaction as a transition from a compact
to a non-compact DS group.

In the present work we realize the above considerations through the
formalism of the VBM.  Our purpose is to examine the various
directions in the ($\l ,\m$)-plane by investigating the respective
changes in the structure of the SU(3) multiplets in terms of model
defined spectroscopic characteristics of rotational nuclei. As such,
we consider here the SU(3) energy splitting and the $g$--$\gamma$
interband transitions, which carry important information about the
link between the two bands.  It is known that the energy splitting of
the multiplet determines to a great extent the systematic behavior of
the SU(3) dynamical symmetry in deformed nuclei \cite{MDRRB97}.

In the VBM relatively simple analytic expressions for the energies and
the transition probabilities can be derived both for the lowest $L=2$
states of any ($\l ,\m$)- multiplet and for all the states of any
($\l , 2$) multiplet. The analytic expressions for the $L=2$ states
allow one to examine the SU(3) characteristics of nuclei in terms of
two-dimensional surfaces in the ($\l ,\m$)-plane, while in the ($\l ,2$)
direction one is able to investigate the behavior of the full set of
states in the multiplet, i.e. the states with $L\geq 2$. On the other
hand, the $L\geq 2$ states of the irreps with $\m > 2$  can be
treated numerically.

In such a way, a relevant combination of analytic and numerical
analyses  could be applied in order to reveal the systematic behavior
of all the states of SU(3) irreps in the ($\l ,\m$)-plane including
the limiting cases of SU(3) group contraction. The collective scheme
of the VBM is constructed by using the irreps with $\l\geq\m$ and
comprises the following two SU(3) contraction limits:

(i) $\l\rightarrow\infty $, with $\m$ finite;

(ii) $\l\rightarrow\infty $, $\m\rightarrow\infty $, with $\m\leq\l$.

Below we provide a detailed study of the most important spectroscopic
characteristics of the $g$ and the $\gamma$ band in the above
limiting cases. It will be shown that our approach gives a reasonable
interpretation of the corresponding experimental data and leads to
rather clear conclusions about the rearrangement of collective
rotational bands in heavy deformed nuclei.

In Sec. II the $g$--$\gamma$ band coupling scheme of the VBM is
briefly presented. In Sec. III we derive analytic expressions for the
energies and the $B(E2)$-transition probabilities for the $2_g$ and
$2_{\gamma}$ states of an arbitrary ($\l ,\m$) multiplet. Using them,
we obtain the analytic behavior of the energy splitting and the
physically meaningful transition ratios in the SU(3) contraction
limits (i) and (ii). In Sec. IV we derive expressions for the
splitting and the transition ratios  for the full set of states
($L\geq 2$) in the ($\l ,2$) multiplets and obtain their analytic
form in the first limiting case ($\l\rightarrow\infty $; $\m =2$). In
Sec. V all analytic results are examined numerically. Also, there we
provide a numerical study of the second limiting case
($\l\rightarrow\infty $, $\m\rightarrow\infty $; $\m\leq\l$) for the
states with $L\geq 2$. The results are discussed together with an
analysis of experimental data. In Sec. VI the conclusions are given.

\section{$g$--$\gamma$ band coupling in the VBM}

The Vector-Boson Model (VBM) with SU(3) dynamical symmetry is founded
on the assumption that the low-lying collective states of deformed
even--even nuclei can be described by means of two distinct kinds of
vector bosons, whose creation operators $\mbox{\boldmath $\xi^{+}$}$
and $\mbox{\boldmath $\eta^{+}$}$ are O(3) vectors and in addition
transform according to two independent SU(3) irreps of the type
$({\l},{\m})=(1,0)$ \cite{p:descr,a:over,p:matr}. The vector bosons
provide a relevant construction of the SU(3) angular momentum and
quadrupole operators like the bosons in the Schwinger realization of
SU(2) \cite{BieLou}. Therefore, they can be considered as natural
building blocks of a model scheme with SU(3) dynamical symmetry.
Also, the vector bosons can be interpreted as quanta of elementary
collective excitations of the nucleus \cite{p:matr}.

In this model an SU(3)-symmetry reducing Hamiltonian is constructed by
using three basic O(3) scalars, which belong to the enveloping
algebra of SU(3) \cite{p:descr}:
\begin{equation}
\label{eq:v}
V=g_{1}L^{2}+g_{2}L\cdot Q\cdot L +g_{3}A^{+}A\ .
\end{equation}
Here $g_{1}$, $g_{2}$ and $g_{3}$ are free parameters; $L$ and $Q$ are
the angular momentum and quadrupole operators respectively; and
$A^{+}=\mbox{\boldmath $\xi^{+}$}^{2}\mbox{\boldmath $\eta^{+}$}^{2}-
(\mbox{\boldmath $\xi^{+}$}\cdot\mbox{\boldmath $\eta^{+}$})^{2}$.

The basis states
\begin{equation}
\label{eq:bast}
\left|\begin{array}{c}({\l},{\m})\\{\a},L,M\end{array}
\right\rangle\ ,
\end{equation}
corresponding to the $SU(3)\supset O(3)$ group reduction, are constructed
by means of the above vector--boson operators and are known as the basis of
Bargmann--Moshinsky \cite{bm:bas,m:bas}.

The quantum number $\a$ in Eq.~(\ref{eq:bast}) distinguishes
the various O(3) irreps, $(L,M)$, appearing in a given SU(3)
irrep $(\l ,\m)$ and labels the different bands of an SU(3)
multiplet. It is an integer number determined through the following
inequality \cite{a:over,m:bas}
\begin{equation}
\max\{ 0,\frac{1}{2}({\m}-L)\}\leq{\a}\leq \min\{ \frac{1}{2}
({\m}-{\beta}),\frac{1}{2}(\l +\m -L-\beta )\}\ ,
\label{eq:minmax}
\end{equation}
where
\begin{equation}
\beta = \left\{ \begin{array}{ll}0, & \mbox{$\l +\m -L$ even} \\
                                 1, & \mbox{$\l +\m -L$ odd}
\end{array} \right.
\nonumber
\end{equation}

In the VBM the $g$- and the lowest $\gamma$- band belong to one and
the same SU(3) multiplet, where $\l$ and $\m$ are even and
$\l\geq\m$. These bands are labeled by two neighboring integer values
of the quantum number $\a$. (More precisely, the states of the $g$-
band are labeled by the largest value of $\a$ appearing in
(\ref{eq:minmax}), while the $\gamma$- band corresponds to the next
smaller $\a$- value.) The so defined multiplet is split with respect
to $\a$.

The above scheme provides a good description of the energy levels and
of the B(E2) transition ratios within and between the $g$- and $\gamma$-
bands \cite{p:descr,MDRRB97}.  The other collective bands, in particular
the lowest $\beta$-band, do not belong to the same irrep.  Therefore,
they are not considered in the framework of this model.

\section{The $L=2$ states in ($\l ,\m$)-plane}

\subsection{\em $L=2$ energy splitting}

Here we consider the $L=2$ energy levels of the $g$- and the
$\gamma$- band in terms of the VBM.  For any ($\l ,\m$) multiplet
($\m\geq 2$), the $2_g$ and $2_{\gamma}$ states are the only possible
ones appearing at angular momentum $L=2$.  They are labeled by the
quantum number $\a$ as follows [See inequality~(\ref{eq:minmax})]:
$\a_{1}={\m}/2-1$ for $2_{\gamma}$ and
$\a_{2}={\m}/2$ for $2_{g}$.
Hence, for the $L=2$ states the Hamiltonian matrix is always
two-dimensional and the corresponding eigenvalue equation has the
form:
\begin{equation}
{\rm det}\left(\begin{array}{cc}
V_{1,1}-\o^{(2)} & V_{1,2} \\
V_{2,1} & V_{2,2}-\o^{(2)}
\end{array}\right) =0 \ ,
\label{eq:sdet}
\end{equation}
where $\o^{(2)}\equiv \o^{L=2}$ are the eigenvalues and
\begin{equation}
V_{j,j'} \equiv \langle\a_{j},2|V|\a_{j'},2\rangle =
\left\langle\begin{array}{c}({\l},{\m})\\{\a}_{j},2,2
\end{array} \right| V
\left|\begin{array}{c}({\l},{\m})\\{\a}_{j'},2,2
\end{array}\right\rangle \ ,
\label{eq:twomat}
\end{equation}
with $j,j'=1,2$, are the corresponding Hamiltonian matrix elements.
We have derived these matrix elements in the form:
\begin{eqnarray}
V_{1,1} &=& \langle (\frac{\m}{2}-1),2|V|(\frac{\m}{2}-1),2\rangle
         =6g_{1}+6g_{2}(2\l +2\m +3)+g_{3}P(\l ,\m ) \ ,
\label{V11}\\
V_{2,2} &=& \langle \frac{\m}{2},2|V|\frac{\m}{2},2\rangle
         =6g_{1}-6g_{2}(2\l +2\m +3)+g_{3}Q(\l ,\m ) \ , \\
V_{1,2} &=& \langle (\frac{\m}{2}-1),2|V|\frac{\m}{2},2\rangle
         =12g_{2}\m -2g_{3}\m (\m -2)\ , \\
V_{2,1} &=& \langle \frac{\m}{2},2|V|(\frac{\m}{2}-1),2\rangle
         =-12g_{2}\l +2g_{3}\l (\l +2\m +2)\ , \label{V21}
\end{eqnarray}
where
\begin{eqnarray}
P(\l ,\m ) &=& \l (\m -2)(\m +2)(\l +2\m +2)+
\m (\m -2)(\m +1)(\m +3)\ ,
\label{Plm} \\
Q(\l ,\m ) &=& \l \m^{2}(\l +2\m +2)+\m (\m -1)(\m +1)(\m +2)\ .
\end{eqnarray}

The energy levels $E_{2}^{g}$ and $E_{2}^{\gamma}$, corresponding to the
$2_g$ and $2_{\gamma}$ states respectively, are determined as
\begin{eqnarray}
E_{2}^{g} &=& \o_{1}^{(2)}-\o^{(0)}\ , \label{E2gsb}\\
E_{2}^{\gamma} &=& \o_{2}^{(2)}-\o^{(0)}\ ,
\label{E2gam}
\end{eqnarray}
where
\begin{equation}
\o_{i}^{(2)}=\frac{1}{2}\left\{ V_{1,1}+V_{2,2}+(-1)^{i}\sqrt{
(V_{1,1}+V_{2,2})^{2}-4(V_{1,1}V_{2,2}-V_{1,2}V_{2,1})}\right\}\ ,
\label{eigenpm}
\end{equation}
$i=1,2$, are the solutions of the eigenvalue equation (\ref{eq:sdet}),
and $\o^{(0)}=g_{3}\m^{2}(\l+\m+1)^{2}$
is the zero-level eigenvalue, corresponding to the ground state $0_g$.
After using Eqs.~(\ref{V11})--(\ref{V21}) we obtain the following
analytic expressions for $E_{2}^{g}$ and $E_{2}^{\gamma}$:
\begin{eqnarray}
E_{2}^{g} &=& 6g_{1}-2Fg_{3}- 2\sqrt{Ag^{2}_{2}+Bg^{2}_{3}-
Cg_{2}g_{3}}\ ,\label{eq:gsb} \\
E_{2}^{\gamma} &=& 6g_{1}-2Fg_{3}+ 2\sqrt{Ag^{2}_{2}+Bg^{2}_{3}-
Cg_{2}g_{3}}\ ,\label{eq:gam}
\end{eqnarray}
where
\begin{eqnarray}
A&=&A({\l},{\m})=9[(2{\l}+2{\m}+3)^{2}-4\l\m]\ ; \label{Alm}\\
B&=&B({\l},{\m})=[{\l}({\l}+2{\m}+2)+{\m}({\m}+1)]^{2}- \nonumber\\
&-&\l\m ({\l}+2{\m}+2)({\m}-2)\ ;\\
C&=&C({\l},{\m})=6(2{\l}+2{\m}+3)[{\l}({\l}+2{\m}+2)+
{\m}({\m}+1)]- \nonumber \\
&-&6\l\m ({\l}+3{\m})\ ; \label{Clm}\\
F&=&F({\l},{\m})={\l}({\l}+2{\m}+2)+2{\m}({\m}+1)\ .
\label{Flm}
\end{eqnarray}

Hence, we derive a model expression for the energy splitting of the SU(3)
multiplet.  It is known that the splitting can be characterized by the
ratio \cite{MDRRB97}:
\begin{equation}
\label{eq:split}
\Delta E_{2}=\frac{E_{2}^{\gamma}-E_{2}^{g}}{E_{2}^{g}}\ .
\end{equation}
In terms of Eqs.~(\ref{eq:gsb}) and (\ref{eq:gam}) the quantity
$\Delta E_{2}$ obtains the following analytic form:
    \begin{equation}
    \Delta E_{2}=\frac{2}{(3g_{1}-Fg_{3})/
    \sqrt{Ag^{2}_{2}+Bg^{2}_{3}-Cg_{2}g_{3}}-1}\ .
    \label{anspl}
    \end{equation}

The expressions, obtained so far, allow us to study analytically the
$g$--$\gamma$ band-mixing interaction and the energy splitting at
$L=2$ in the ($\l ,\m$)-plane. In particular we are able to reproduce
analytically the SU(3) contraction limits:

(i) $\l\rightarrow\infty $, with $\m$ finite;

(ii) $\l\rightarrow\infty $, $\m\rightarrow\infty $, with $\m\leq\l$.
Since the difference $\l -\m$ is always finite, we take for
definiteness $\m =\l$.

In each of these limits we estimate the $\l$- and/or $\m$- dependence
of the matrix elements (\ref{V11})--(\ref{V21}), as well as the
analytic behavior of the splitting ratio $\Delta E_{2}$.

In case (i) the matrix elements are determined by the corresponding
highest degrees of $\l$.  Thus for $\m >2$ the Hamiltonian matrix
$(V_{i,j})$ obtains the following asymptotic form:
\begin{equation}
(V)_{\l\rightarrow\infty}\sim
\left(\begin{array}{cc}\l^{2} & {*}\\ \l^{2} &
\l^{2}\end{array}\right)\ ,
\label{Vlim1}
\end{equation}
where the upper off-diagonal element (denoted by ${*}$) does not
depend on $\l$.  Then the relative contribution of the off-diagonal
(band-mixing) terms in the eigenvalue equation (\ref{eq:sdet})
decreases with the increase of $\l$ as $\l^{2}/\l^{4}=1/\l^{2}$.  For
$\m= 2$ the term $V_{1,1}$ is proportional to $\l$ instead of
$\l^{2}$ [See Eqs.~(\ref{V11}) and (\ref{Plm})], so that in this
particular case the off-diagonal contribution decreases as $1/\l$.

In the same limiting case the functions (\ref{Alm})--(\ref{Flm})
have the following asymptotic behavior:
    $$A_{\l\rightarrow\infty}=36\l^{2};\ \ \
      B_{\l\rightarrow\infty}=\l^{4};\ \ \
      C_{\l\rightarrow\infty}=12\l^{3};\ \ \
      F_{\l\rightarrow\infty}=\l^{2} \ .$$
After applying them in Eq.~(\ref{anspl}), we find the
analytic limit of the splitting ratio (\ref{anspl}):
\begin{equation}
\lim_{\l\rightarrow\infty}\Delta E_{2}=
\frac{2}{-g_{3}/|g_{3}|-1}\ .
\label{lime}
\end{equation}
We remark that the application of the VBM in rare earth nuclei and
actinides requires $g_{3}<0$ \cite{MDRRB97}, which gives in
(\ref{lime})
    \begin{equation}
    \lim_{\l\rightarrow\infty}\Delta E_{2}=\infty.
    \label{lde1}
    \end{equation}
Therefore, in this case the SU(3)-multiplet is completely split.

Consider now the limiting case (ii), $\l = \m \rightarrow\infty $.
Then the asymptotic form of the matrix $(V_{i,j})$ is:
    \begin{equation}
    (V)_{\l =\m\rightarrow\infty}\sim
    \left(\begin{array}{cc}\l^{4} & \l^{2}\\ \l^{2} & \l^{4}
    \end{array}\right)\ .
    \label{Vlim2}
    \end{equation}
Here we find that the relative magnitude of the band-mixing
interaction decreases as $\l^{4}/\l^{8}=1/\l^{4}$, i.e., more rapidly
in comparison to the previous case.

Furthermore, in the limiting case (ii) one has:
    $$A_{\l =\m\rightarrow\infty}=108\l^{2};\ \ \
      B_{\l =\m\rightarrow\infty}=13\l^{4};\ \ \
      C_{\l =\m\rightarrow\infty}=72\l^{3};\ \ \
      F_{\l =\m\rightarrow\infty}=5\l^{2} \ .$$
Then the SU(3) splitting ratio goes to:
    \begin{equation}
    \lim_{\l =\m\rightarrow\infty}\Delta E_{2}=
    \frac{2}{-(5/\sqrt{13})g_{3}/|g_{3}|-1}\ .
    \end{equation}
For $g_{3}<0$ we obtain
    \begin{equation}
    \lim_{\l =\m\rightarrow\infty}\Delta E_{2}=
    2/(5/\sqrt{13}-1)=5.17\ .
    \label{lde2}
    \end{equation}
Therefore, in this case the band-mixing interaction
vanishes, while the energy splitting between the two bands
remains finite.

\subsection{\em Transition ratios in the $L=2$ states}

Here we turn to the electromagnetic transition probabilities for the states
$2_{g}$, $2_{\gamma}$ and $0_{g}$.  In particular it is of interest to
consider the following B(E2) transition ratios:
\begin{eqnarray}
R_{1}(2)&=&
\frac{B(E2;2_\gamma \rightarrow 2_g)}{B(E2;2_g\rightarrow 0_g)}\ ;
\label{R1E2}\\
R_{2}(2)&=&
\frac{B(E2;2_\gamma \rightarrow 2_g)}{B(E2;2_\gamma\rightarrow
0_g)}\ .
\label{R2E2}
\end{eqnarray}
The first of them, $R_{1}(2)$, gives the relative magnitude of the
$g$--$\gamma$ interband transition probability with respect to the ground
intraband one.  Thus it naturally characterizes the link between the two
bands within the multiplet.  The second ratio represents one of the widely
used collective characteristics of nuclei related to Alaga rules.  Both
quantities (\ref{R1E2}) and (\ref{R2E2}) can be obtained from the
experimental data on deformed nuclei and therefore have a direct physical
meaning.

In order to derive analytic expressions for the above ratios we
calculate the matrix elements of the quadrupole operator $Q_{0}$
between the eigenstates
\begin{eqnarray}
\left.|\omega^{(2)}_{i}\right\rangle &=&
C^{{(2)}}_{i1}
\left|\begin{array}{c}({\l},{\m})\\{\m}/2-
1,2,2\end{array}\right\rangle +C^{(2)}_{i2}
\left|\begin{array}{c}({\l},{\m})\\{\m}/2,2,2\end{array}
\right\rangle\ , \qquad i=1,2;\\
\left.|\omega^{(0)}\right\rangle &=& C^{(0)}
\left|\begin{array}{c}({\l},{\m})\\{\m}/2,0,0\end{array}
\right\rangle\ ,
\end{eqnarray}
of the VBM Hamiltonian (\ref{eq:v}). (It should be remembered that
the eigenvalues $\omega^{(2)}_{1}$, $\omega^{(2)}_{2}$ and
$\omega^{(0)}$ correspond to the $2_{g}$, $2_{\gamma}$ and $0_{g}$
states respectively.) After applying analytically the formalism developed in \cite{MDRRB97} we
obtain the following matrix elements:
\begin{eqnarray}
\left\langle\omega^{(2)}_{1}\right|Q_{0}
|\left.\omega^{(2)}_{2}\right\rangle
&=& \frac{4}{7}\frac{{\l}(C^{(2)}_{21})^2+{\m}(C^{(2)}_{22})^2+
(2{\l}+2{\m}+3)C^{(2)}_{21}C^{(2)}_{22}}{C^{(2)}_{11}C^{(2)}_{22}-
C^{(2)}_{21}C^{(2)}_{12}} \ ; \\
\left\langle\omega^{(0)}_{1}\right|Q_{0}
|\left.\omega^{(2)}_{1}\right\rangle
&=& \sqrt{6} C^{(0)} \frac{{\m}C^{(2)}_{22}-{\l}C^{(2)}_{21}}
{C^{(2)}_{11}C^{(2)}_{22}-C^{(2)}_{12}C^{(2)}_{21}}\ ; \\
\left\langle\omega^{(0)}\right|Q_{0}|\left.\omega^{(2)}_{2}\right
\rangle
&=& \sqrt{6} C^{(0)} \frac{{\l}C^{(2)}_{11}-{\m}C^{(2)}_{12}}
{C^{(2)}_{11}C^{(2)}_{22}-C^{(2)}_{12}C^{(2)}_{21}}\ .
\end{eqnarray}
The wave-function coefficients are determined as
\begin{eqnarray}
C^{(2)}_{i1}&=&(f^{(2)}_{11}+2h_{i2}f^{(2)}_{21}+
h^{2}_{i2}f^{(2)}_{22})^{-\frac{1}{2}}\ ;\\
C^{(2)}_{i2}&=&h_{i2}C^{(2)}_{i1}\ ,\qquad i=1,2 \\
C^{(0)}&=&(f^{(0)})^{-\frac{1}{2}}\ .
\end{eqnarray}
Here
    \begin{eqnarray}
    f^{(2)}_{11}=\left\langle\begin{array}{c}({\l},{\m})\\{\m}/2-1,2
    \end{array} \right|
    \left. \begin{array}{c}({\l},{\m})\\{\m}/2-1,2
    \end{array}\right\rangle =
    \frac{1}{30}R({\l},{\m})
    \sum_{l=0}^{{\m}-2}\left(\begin{array}{c}{\m}/2\\l/2
    \end{array}\right)
    S^{l}({\l},{\m})\frac{(l+1)({\m}-l)}{{\m}^2
    (\lambda +l+6)}\nonumber \\
    \times [({\m}-l)({\l}+2)({\l}+3)({\l}+5)-{\m}{\l}({\l}+4)
    ({\l}+l+6)]\ ;
    \label{f11}
    \end{eqnarray}

    \begin{eqnarray}
    f^{(2)}_{21}&=&\left\langle\begin{array}{c}({\l},{\m})\\{\m}/2,2
    \end{array} \right|
    \left. \begin{array}{c}({\l},{\m})\\{\m}/2-1,2
    \end{array}\right\rangle=
    \frac{1}{15}R({\l},{\m})
    \sum_{l=0}^{{\m}-2}\left(\begin{array}{c}{\m}/2\\l/2
    \end{array}\right)
    S^{l}({\l},{\m})\frac{(l+1)({\m}-l)}{{\m}}\ ;\ \ \ \
    \end{eqnarray}

    \begin{eqnarray}
    f^{(2)}_{22}&=&\left\langle\begin{array}{c}({\l},{\m})\\{\m}/2,2
    \end{array} \right|
    \left. \begin{array}{c}({\l},{\m})\\{\m}/2,2
    \end{array}\right\rangle=
    \frac{1}{15}R({\l},{\m})
    \sum_{l=0}^{{\m}}\left(\begin{array}{c}{\m}/2\\l/2
    \end{array}\right)
    S^{l}({\l},{\m})(l+1)(l+2)\ ; \ \ \ \ \ \
    \end{eqnarray}

    \begin{eqnarray}
    f^{(0)}&=&\left\langle\begin{array}{c}({\l},{\m})\\{\m}/2,0
    \end{array} \right|
    \left. \begin{array}{c}({\l},{\m})\\{\m}/2,0
    \end{array}\right\rangle=
    R({\l},{\m})
    \sum_{l=0}^{{\m}}\left(\begin{array}{c}{\m}/2\\l/2
    \end{array}\right)S^{l}({\l},{\m})
    \frac{({\l}+{\m}-l)({\l}+l+4)}
    {{\m}({\l}+3)({\l}+{\m}+4)} \ , \ \ \ \
    \label{f0}
    \end{eqnarray}
are the corresponding overlap integrals obtained by the general
expression in \cite{a:over}, with
    \begin{eqnarray}
    R({\l},{\m})&=&({\l}+3)!!({\m}!!)^2\ ,\\
    S^{l}({\l},{\m})&=&((l-1)!!)^{2}\frac{({\l}+
    {\m}-l-2)!!({\l}+{\m}+4)!!}
    {({\l}+l+4)!!}.
    \end{eqnarray}
In addition [see Eqs. (13)--(15) in ref. \cite{MDRRB97}]
    \begin{eqnarray}
    h_{i2}=-\frac {(V_{11}-\omega^{(2)}_{i})}{V_{12}}=\left[-3g_{2}
    (2{\l}+2{\m}+3)+g_{3}[({\l}+{\m})^2+2{\l}+{\m}]+ \right.
    \nonumber \\
    \left.(-1)^{i}\sqrt {A({\l},{\m})g^{2}_{2}+B({\l},{\m})g^{2}_{3}-
    C({\l},{\m})g_{2}g_{3}}\right] /(6g_{2}{\m}-g_{3}{\m}({\m}-2))\ ,
    \label{hi2}
    \end{eqnarray}
with $A({\l},{\m})$, $B({\l},{\m})$ and
$C({\l},{\m})$ being defined in Eqs.~(\ref{Alm})--(\ref{Clm}).

By using the general expression for the B(E2) transition
probability between two of the above eigenstates
    \begin{eqnarray}
    B(E2;L_\nu \rightarrow L'_{\nu''}) &=&
    \left(\begin{array}{ccc}
    L' & 2 & L \\
    -L& 0 & L\end{array}\right)^{-2}
    \left| \left\langle\omega^{(L')}_{\nu''}\right|Q_{0}
    |\left.\omega^{(L)}_{\nu}
    \right\rangle \right|^{2} \
    \end{eqnarray}
    ($L,L'=0,2$; $\nu ,{\nu}''=g,\gamma$),
we have studied analytically the transition ratios (\ref{R1E2}) and
(\ref{R2E2}) in the two limiting cases considered in the previous
subsection.

We have analyzed the explicit expressions for the overlap integrals
(\ref{f11})--(\ref{f0}) and the $h_{i2}$-factors (\ref{hi2}).  In
this way we have deduced that in both limits, (i) and (ii), the
overlap integrals increase to infinity.  On the other hand one can
verify that this behavior is compensated consistently in the ratios
(\ref{R1E2}) and (\ref{R2E2}), where the total contribution of the
integrals and the $h_{i2}$-factors is finite.

Thus, for the case (i) ($\l\rightarrow\infty $, with $\m$ finite) we
have obtained the following analytic limits of the transition ratios
$R_{1}(2)$ and $R_{2}(2)$:
    \begin{eqnarray}
    \lim_{\lambda \rightarrow \infty} \frac{B(E2;2_\gamma
    \rightarrow 2_g)}
    {B(E2;2_g\rightarrow 0_g)} &=& 0 \ ;\\
    \label{r1lim1}
    \lim_{\lambda \rightarrow \infty} \frac{B(E2;2_\gamma
    \rightarrow 2_g)}
    {B(E2;2_\gamma \rightarrow 0_g)} &=& \frac{10}{7} \left
    (\frac{\mu +2}{2\mu}\right )^{2}\ .
    \label{r2lim1}
    \end{eqnarray}

So, in this case we find that the relative magnitude of the
$g$--$\gamma$ interband transition is zero, while the ratio
$R_{2}(2)$ obtains finite values depending on the quantum number
$\mu$.  We remark that for $\mu =2$ one has $R_{2}(2)=10/7$, which
is the standard Alaga value.

In case (ii) ($\mu =\lambda \rightarrow \infty$) we obtain the
following limits:
\begin{eqnarray}
\lim_{\lambda = \mu \rightarrow \infty} \frac{B(E2;2_\gamma
\rightarrow 2_g)}
{B(E2;2_g\rightarrow 0_{g})} &=& \frac{10}{7} \frac
{(c_{2}^{2}+4c_{2}+1)^2}
{(c_{2}^{2}+c_{2}+1)(c_{2}-1)^2} \approx  0.172\ ; \label{r1lim2} \\
\lim_{\lambda = \mu \rightarrow \infty} \frac{B(E2;2_{\gamma}
\rightarrow 2_g)}
{B(E2;2_{\gamma}\rightarrow 0_{g})} &=& \frac{10}{7}
\frac {(c_{2}^{2}+4c_{2}+1)^2 (c_{1}^{2}+c_{1}+1)}
{(c_{2}^{2}+c_{2}+1)^{2}(c_{1}-1)^2} \approx  0.304 \ , \label{r2lim2}
\end{eqnarray}
with $c_{1}=-4-\sqrt{13}\approx -7.606$ and
$c_{2}=-4+\sqrt{13}\approx -0.394$.

In this case one finds that both ratios, $R_{1}(2)$ and
$R_{2}(2)$, remain finite.

We remark that all obtained limits do not depend on the model
parameters.  (It is assumed that $g_{1}$, $g_{2}$, and $g_{3}$ are
finite, with $g_{2}<0$ and $g_{3}<0$.)

\section{The ($\l , 2$)-direction}
\subsection{\em SU(3) splitting in $L\geq 2$ states}

For the  $(\l ,2)$- irreps the $g$- and the $\gamma$-bands are the
only possible ones appearing in the corresponding SU(3) multiplets.
They are labeled by the quantum numbers $\a_{2} =1$ and $\a_{1} =0$
respectively [See inequality~(\ref{eq:minmax})]. In the even angular
momentum states the  Hamiltonian matrix is always two-dimensional,
while for the odd states of the $\gamma$- band one has a single
matrix element. Hence for the $(\l ,2)$- multiplets one is able to
derive analytic expressions for the spectroscopic characteristics of
the {\em full set of states} ($L\geq 2$) in a way similar to that of the
previous section. That is why we do not explain in detail all steps
of analytic calculations and report only the final results in this
direction.

So, for a given $(\l ,2)$ multiplet the energy levels $E^{g}(L)$ and
$E^{\gamma}(L)$ of the $g$ and the $\gamma$-band can be written in the
following form:
\begin{eqnarray}
E^{g}(L)&=& \tilde{B}+\tilde{A}L(L+1)-\mod{\tilde{B}} R^{(L)} \ ,
\label{Eground}  \\
E^{\gamma}(L_{even})&=& \tilde{B}+\tilde{A}L(L+1)+
\mod{\tilde{B}}R^{(L)} \ ;\label{Egammae}  \\
E^{\gamma}(L_{odd})&=&2\tilde{B}+(\tilde{A}+g_{3})L(L+1) \ ,
\label{Egammao}
\end{eqnarray}
where
\begin{eqnarray}
\tilde{A}&=&\tilde{A}(g_1,g_2,g_3)=g_{1}-(2\l +5)g_{2} -g_3, \\
\tilde{B}&=&\tilde{B}({\l},g_2,g_3)=6(2\l +5)g_{2}-
2(\l +3)^{2} g_{3},
\end{eqnarray}
and
\begin{equation}
R^{(L)}=\sqrt{1+aL(L+1)+bL^{2}(L+1)^{2}}\ .
\end{equation}
with
\begin{eqnarray}
a=a({\l},g_2,g_3)&=& -\frac{4}{\tilde{B}^{2}}
\left\{(\l+3)[(\l+3)g_{3}-
6g_{2}]g_{3}-3(g_{3}-6g_{2})g_{2}\right\}, \\
b=b({\l},g_2,g_3)&=&\frac{1}{\tilde{B}^{2}}
\left(g_{3}-6g_{2}\right)^{2}\ .
\end{eqnarray}

Now we introduce the following energy ratio
\begin{equation}
\label{eq:splitL}
\Delta E_{L}=\frac{E_{L}^{\gamma}-E_{L}^{g}}{E_{2}^{g}}\ ,
\end{equation}
which is more general compared to Eq.(\ref{eq:split}) and
characterizes the magnitude of the energy splitting in any even
angular momentum state of a given SU(3) multiplet.

By using Eqs.(\ref{Eground}) and (\ref{Egammae}) we obtain
$\Delta E_{L}$ in the following analytic form
\begin{equation}
\Delta E_{L}=\frac{2\mod{\tilde{B}}R^{(L)}}
{6\tilde{A}-\mod{\tilde{B}}R^{(2)}+\tilde{B}} \ ,
\label{ELratio}
\end{equation}
which in the SU(3) contraction limit goes to
\begin{equation}
\lim_{\stackrel{\l\rightarrow\infty}{\m =2}}\Delta E_{L}=
\frac{2}{-g_{3}/|g_{3}|-1}\ .
\label{limeL}
\end{equation}

For $g_{3}<0$ one has
\begin{equation}
\lim_{\stackrel{\l\rightarrow\infty}{\m =2}}\Delta E_{L}=\infty.
\label{lde1L}
\end{equation}

Thus we find that for all even states of a given $(\l ,2)$-
multiplet the SU(3) splitting goes to infinity in the same way [see
also Eqs.(\ref{lime}) and (\ref{lde1})].

\subsection{\em Transition ratios in the ($\l , 2$)-direction}

For the ($\l , 2$)-direction the B(E2) transitions between the states
of a given multiplet can be examined through the following [more
general compared to (\ref{R1E2}) and (\ref{R2E2})] transition ratios:
\begin{eqnarray}
R_1(L) &=& \frac{B(E2;L_\gamma \rightarrow L_g)}
{B(E2;L_g\rightarrow (L-2)_g)} \ , \qquad\qquad L=even \ ,
\label{R1(L)} \\
R_2(L) &=& \frac{B(E2;L_\gamma\rightarrow L_g)}
{B(E2;L_\gamma\rightarrow (L-2)_g)}\ , \qquad\qquad L=even \ ,
\label{R2(L)} \\
R_3(L) &=& \frac{B(E2;L_\gamma \rightarrow (L+1)_g)}
{B(E2;L_\gamma\rightarrow (L-1)_g)} \ , \qquad\qquad L=odd \ .
\label{R3(L)}
\end{eqnarray}

The first two ratios, $R_1(L)$ and $R_2(L)$, have the same physical
meaning as the ratios (\ref{R1E2}) and (\ref{R2E2}) of the previous
section. The third ratio, $R_3(L)$, involves the odd angular momentum
states in the study. In such a way we investigate the transition
characteristics of the full set of states in a given SU(3) multiplet.

In the case of a ($\l , 2$)-multiplet the Hamiltonian eigenstates are
constructed as
\begin{eqnarray}
\left.|\omega^{(L)}_{i}\right\rangle &=&
C^{{(L)}}_{i1}
\left|\begin{array}{c}({\l},2)\\
0,L\end{array}\right\rangle +C^{(L)}_{i2}
\left|\begin{array}{c}({\l},2)\\1,L\end{array}
\right\rangle\ , \qquad i=1,2,\ L=even;\\
\left.|\omega^{(L)}_{odd}\right\rangle &=& C^{(L)}_{odd}
\left|\begin{array}{c}({\l},2)\\ 0,L\end{array}
\right\rangle\ , \qquad L=odd.
\end{eqnarray}

The necessary transition matrix elements are derived in the form
\begin{eqnarray}
\left\langle\omega^{(L)}_{1}\right|Q_{0}
|\left.\omega^{(L)}_{2}\right\rangle
&=&\frac{12}{(L+1)(2L+3)}
\left [ ({\l}+2-L)(C^{(L)}_{21})^2 \right. \nonumber\\
&+&[2{\l}+5+L(L-1)]
C^{(L)}_{21}C^{(L)}_{22} \nonumber\\
&+&\left. L(L-1)(C^{(L)}_{22})^2 \right ]/
\left [C^{(L)}_{11}C^{(L)}_{22}-C^{(L)}_{21}C^{(L)}_{12}
\right ]\ ;
\label{omega1LL}
\end{eqnarray}
\begin{eqnarray}
\left\langle\omega^{(L-2)}_{1}\right|Q_{0}
|\left.\omega^{(L)}_{1}\right\rangle
&=&\frac{6}{\sqrt{L(2L-1)}}
\left [({\l}+4-L)C^{(L-2)}_{11}C^{(L)}_{22}\right. \nonumber\\
&-&({\l}+2-L)C^{(L-2)}_{12}C^{(L)}_{21} \nonumber\\
&+&\left. 2C^{(L-2)}_{12}C^{(L)}_{22}\right ]/
\left [C^{(L)}_{11}C^{(L)}_{22}-C^{(L)}_{21}C^{(L)}_{12}
\right ]\ ;
\label{omega2LL}
 \end{eqnarray}
\begin{eqnarray}
\left\langle\omega^{(L-2)}_{1}\right|Q_{0}
|\left.\omega^{(L)}_{2}\right\rangle
&=&\frac{-6}{\sqrt{L(2L-1)}}
\left [({\l}+4-L)C^{(L-2)}_{11}C^{(L)}_{12} \right.\nonumber\\
&-&({\l}+2-L)C^{(L-2)}_{12}C^{(L)}_{11}\nonumber\\
&+&\left. 2C^{(L-2)}_{12}C^{(L)}_{12} \right ]/
\left [C^{(L)}_{11}C^{(L)}_{22}-C^{(L)}_{21}C^{(L)}_{12}
\right ] \ ;
\label{omega3LL}
\end{eqnarray}
\begin{eqnarray}
\left\langle\omega^{(L+1)}_{1}\right|Q_{0}
|\left.\omega^{(L)}_{2}\right\rangle
&=& -\frac{6}{(L+2)\sqrt{L+1}}C^{(L)}_{\rm odd}
\left [ (2{\l}-L+4)C^{(L+1)}_{22} \right.\nonumber\\
&+&\left. ({\l}-L+1)C^{(L+1)}_{21} \right ]/
\left [C^{(L+1)}_{11}C^{(L+1)}_{22}-
C^{(L+1)}_{21}C^{(L+1)}_{12}\right ] \ ;
\label{omega1odd}
\end{eqnarray}
\begin{eqnarray}
\left\langle\omega^{(L-1)}_{1}\right|Q_{0}
|\left.\omega^{(L)}_{2}\right\rangle
&=& -\frac{12}{(L+1)\sqrt{L}}\frac{1}{C^{(L)}_{\rm odd}}
\left [({\l}-L+3)C^{(L-1)}_{11}-(L-1)C^{(L-1)}_{12}\right ] \ ,
\label{omega2odd}
\end{eqnarray}
with the  wave-function coefficients
\begin{eqnarray}
C^{(L)}_{i1}&=&\left (f^{(L)}_{11}+2h^{(L)}_{i2}f^{(L)}_{21}+
(h^{(L)})^{2}_{i2}f^{(L)}_{22}\right )^{-\frac{1}{2}}\ ,
\qquad L=even \ ;\\
C^{(L)}_{i2}&=&h^{(L)}_{i2}C^{(L)}_{i1}\ ,
\qquad L=even \ , i=1,2 \\
C^{(L)}_{odd}&=&(f^{(L)}_{odd})^{-\frac{1}{2}}\ ,
\qquad L=odd \ .
\end{eqnarray}

The corresponding overlap integrals are obtained in the form
\begin{eqnarray}
        f^{(L)}_{11}&=&\left\langle\begin{array}{c}({\l},{2})\\ 0,L
    \end{array} \right|
    \left. \begin{array}{c}({\l},2)\\ 0,L
    \end{array}\right\rangle =
    S^{L}({\l})\frac{[(L^2+L+1){\l}+L^3+4L^2+2L+2]}{L(L-1)}\ ;
    \label{fL11}
    \end{eqnarray}
    \begin{eqnarray}
    f^{(L)}_{21}&=&\left\langle\begin{array}{c}({\l},2)\\ 1,L
    \end{array} \right|
    \left. \begin{array}{c}({\l},2)\\ 0,L
    \end{array}\right\rangle=
    S^{L}({\l})({\l}+L+4)\ ;
    \label{fL21}
    \end{eqnarray}
    \begin{eqnarray}
    f^{(L)}_{22}&=& \left\langle\begin{array}{c}({\l},2)\\ 1,L
    \end{array} \right|
    \left. \begin{array}{c}({\l},2)\\ 1,L
    \end{array}\right\rangle=
    S^{L}({\l})\frac{[2({\l}-L+2)({\l}+L+4)+(L+1)(L+2)]}{({\l}-L+2)}\
    ; \label{fL22}
 \end{eqnarray}
    \begin{eqnarray}
     f^{(L)}_{\rm odd}&=& \left\langle\begin{array}{c}({\l},2)\\ 0,L
    \end{array} \right|
    \left. \begin{array}{c}({\l},2)\\ 0,L
    \end{array}\right\rangle=
     S^{L}_{\rm odd}({\l})\frac{(L+1)(L+2)({\l}+2)}{2L(L-1)}\ ,
    	\label{fLodd}
    \end{eqnarray}
   with
\begin{eqnarray}
S^{L}({\l})&=&\frac{2L!({\l}-L+2)!!({\l}+L+1)!!}{(2L+1)!!}
\nonumber \\
S^{L}_{\rm odd}({\l})&=&\frac{2L!({\l}-L+1)!!({\l}+L+2)!!}
{(2L+1)!!}\ .
\label{SL}
\end{eqnarray}

Also we have
\begin{eqnarray}
h_{i2}^{(L)}&=&\left [-6g_2[(2{\l}+5)+L(L-1)]+
g_3[2({\l}+3)^2-L(L+1)]\right. \nonumber \\
&+&\left. (-1)^{i}\tilde{B}R^{(L)}\right ]/
[12g_2L(L-1)] \ .
\end{eqnarray}

Eqs.~(\ref{fL11})--(\ref{fLodd}) show that the overlap integrals
increase to infinity with the increase of the quantum number $\l$.
However, similarly to the previous section, one can verify that this
behavior is compensated consistently in the ratios
(\ref{R1(L)})--(\ref{R3(L)}).

After using the above analytic form of the matrix elements
(\ref{omega1LL})--(\ref{omega2odd}) we have obtained the SU(3)
contraction limits of the ratios $R_1(L)$, $R_2(L)$ and $R_3(L)$
[Eqs. (\ref{R1(L)})--(\ref{R3(L)})]:
\begin{eqnarray}
\lim_{\stackrel{\l\rightarrow\infty}{\m =2}}
\frac{B(E2;L_\gamma \rightarrow L_g)}
{B(E2;L_g\rightarrow (L-2)_g)}&=&0 \ ;
\label{limB1L} \\
\lim_{\stackrel{\l\rightarrow\infty}{\m =2}}
\frac{B(E2;L_\gamma \rightarrow L_g)}
{B(E2;L_\gamma\rightarrow (L-2)_g)}&=&6\frac{(L-1)(2L+1)}
{(L+1)(2L+3)} \ ;
\label{limB2L}  \\
\lim_{\stackrel{\l\rightarrow\infty}{\m =2}}
\frac{B(E2;L_\gamma \rightarrow (L+1)_g)}
{B(E2;L_\gamma\rightarrow (L-1)_g)} &=&\frac{(L-1)}{(L+2)}\ .
\label{limB3L}
\end{eqnarray}

Thus, we find that for all the states ($L\geq 2$) of the Hamiltonian
the relative magnitude of the $g$--$\gamma$ interband transitions
goes to zero. Also, we see that the ratios $R_{2}(L)$ and $R_{3}(L)$
go to the corresponding standard Alaga rules.

\section{Results and Discussions}

The theoretical results given above allow one to examine the
mechanism of the SU(3) symmetry reduction in the space of the
$(\l ,\m )$ irreps as well as to identify its manifestation in
reference to the experimental data on heavy deformed nuclei.

The analytic study of the Hamiltonian matrix elements shows
[Eqs.~(\ref{Vlim1}) and (\ref{Vlim2})] how the increase in the
quantum numbers $\l$ and/or $\m$ is connected with the corresponding
decrease in the $g$--$\gamma$ band-mixing interaction within the
framework of the SU(3) symmetry.  Generally this result illustrates
the behavior of the energy-mixing in the $(\l ,\m )$-plane.  In both
limits, (i) and (ii), the $g$--$\gamma$ mixing decreases
asymptotically to zero. Similar limiting behavior of the $L\geq 2$
matrix elements in $(\l , 2 )$-direction has been established in our
previous work (See Sec. IV--C of Ref.~\cite{MDRRB97}). Thus in all
limiting cases the SU(3) symmetry disappears completely and the two
bands do not belong to the same SU(3) multiplet anymore.

It is appropriate at this point to elucidate the meaning of the above
consideration in terms of the SU(3) group contraction process
\cite{CBVR86}--\cite{JPD93}. This
process corresponds to a renormalization of the quadrupole
operator, $Q\leftarrow Q/\sqrt{\langle C_{2}\rangle}$, with
    \begin{equation}
    \langle C_{2}\rangle = (\l +2\m )(\l +2\m +6)+3\l (\l +2 )
    \label{C2}
    \end{equation}
being the eigenvalue of the second order Casimir operator of SU(3).  The
following commutation relations between the angular momentum and the
renormalized quadrupole operators are then valid:
\begin{eqnarray}
{[L_{m},L_{n}]} &=& -\sqrt{2}C_{1m1n}^{1m+n}L_{m+n} \label{LL}, \\
{[L_{m},Q_{n}]} &=& \sqrt{6}C_{1m2n}^{2m+n}Q_{m+n} \label{LQ}, \\
{[Q_{m},Q_{n}]} &=& 3\sqrt{10}C_{2m2n}^{1m+n}\frac{L_{m+n}}
{\langle C_{2}\rangle} \ . \label{QQ}
\end{eqnarray}
They differ from the standard SU(3) commutation relations by the factor
$\langle C_{2}\rangle$ in the right-hand side of (\ref{QQ}).  Taking into
account Eq.~(\ref{C2}), one finds that in both limits (i) and (ii),
considered in the present work, the commutator (\ref{QQ}) vanishes and the
commutation relations of the algebra of the triaxial rotor group
T$_{5}\wedge$SO(3) hold.  In such a way the vanishing $g$--$\gamma$
band-mixing could be interpreted as a transition from a compact to a
non-compact DS group.

Let us now analyze the behavior of the splitting and transition
ratios of Secs. 3 and 4 in the $(\l ,\m )$-plane. For this purpose we
use the analytic expressions for numerical calculations. In the
particular case of $L\geq 2$ states in ($\mu =\lambda \rightarrow
\infty$) direction, which is not accessible analytically, we apply
numerically the algorithm developed in Ref.~\cite{MDRRB97}. All
calculations are carried out for the same set of fixed model parameters
$g_{1}=1$, $g_{2}=-0.2$, $g_{3}=-0.25$. These values belong to the
corresponding parameter regions obtained for a group of rare earth
nuclei and actinides [See Table~2 in Ref.~\cite{MDRRB97}]. In this
respect they can be considered as an overall set of model parameters.
Also, it should be emphasized that in the SU(3) contraction limiting
cases the various sets of (finite) parameter values give the same
asymptotic behavior of the model quantities.

In Fig.~1 the splitting ratio $\Delta E_{2}$ [Eq.~(\ref{anspl})] is
plotted as a function of the quantum numbers $\l$ and $\m$. In the
limiting case (i) ($\l\rightarrow\infty $, with $\m$ finite) the
two-dimensional surface shows a rapid increase of $\Delta E_{2}$,
while in case (ii) ($\mu =\lambda \rightarrow \infty$) the splitting
ratio gradually saturates towards the constant value $\sim 5.17$ [See
Eq.~(\ref{lde2})]. In Fig.~2 the splitting ratio $\Delta E_{L}$
[Eq.~(\ref{eq:splitL})] is plotted as a function of the quantum
number $\l$  for $L=2,4\dots 12$. In the case $\l\rightarrow\infty $,
$\m =2$ the energy splitting goes to infinity with almost equal
values for all angular momenta (Fig. 2(a)). In case (ii)  $\Delta
E_{L}$ trends to finite values which increase with $L$ (Fig. 2(b)).
So, in the first limiting case the complete reduction of the SU(3)
symmetry leads to a large energy separation between the bands in the
multiplet, while in the second case (for finite angular momenta) the
bands remain close to each other, but their mutual disposition does
not depend on the Hamiltonian parameters anymore, so that it should
not be associated with any band coupling.

The experimental $\Delta E_{2}$ ratios of several rare earth nuclei
and actinides are given in Table~1.  They vary within the limits
$5\leq\Delta E_{2}\leq 20$, for the rare earths and $13\leq\Delta
E_{2}\leq 25$, for the actinides. The behavior of the splitting
ratios is clear:  The $\Delta E_{2}$ ratio generally increases
towards the middle of the rotational region.  This is illustrated in
Table~1 through the number of the nucleon pairs (or holes) in the
valence shells, $N$.  (The number $N$ is a well established
characteristic of nuclear collectivity used in the IBM \cite{IA}.) A
clearly pronounced increase of $\Delta E_{2}$ with increasing $N$ is
observed for the isotopes of Sm, Gd, Er, Yb, and W. Similar behavior
of the energy splitting is observed in the $L>2$ states of these
nuclei \cite{Sakai}.  One concludes that the data show that the SU(3)
splitting increases toward the midshell regions.

We turn now to the analysis of the interband transition ratios. In
Fig.~3 the theoretical ratio $R_{1}(2)$, Eq.~(\ref{R1E2}), is plotted
as a two-dimensional function of the quantum numbers $\l$ and $\m$.
In the limiting case (i) ($\l\rightarrow\infty $, with $\m$ finite)
the $R_{1}$ surface shows a rapid decrease to zero. In case (ii)
($\mu =\lambda \rightarrow \infty$) $R_{1}(2)$ decreases gradually
and saturates towards the constant value $\sim 0.172$ [See
Eq.~(\ref{r1lim2})]. In Fig.~4 the transition ratio $R_{1}(L)$
[Eq.~(\ref{R1(L)})] is plotted as a function of the quantum number
$\l$ for $L=2,4\dots 12$. In the case $\l\rightarrow\infty $, $\m =2$
it goes to zero with almost equal values for all angular momenta
(Fig. 4(a)). In case (ii) $R_{1}(L)$ obtains finite values which
decrease with $L$ (Fig. 4(b)). Thus in the first limiting case the
$g$--$\gamma$ interband transition link vanishes rapidly, while in
the second case (for finite angular momenta) the relative magnitude
of the interband transition probability remains non-zero.  However,
as in the energy splitting, the $R_{1}$-ratios do not depend on the
Hamiltonian parameters anymore. Therefore, they should not be treated
in terms of the SU(3) symmetry anymore.

The experimental $R_{1}(2)$-values for several rare earth nuclei and
actinides are given in Table~1.  Here one observes a rather
spectacular decrease of $R_{1}(2)$ towards the midshell regions.  The
best examples (with the largest number of available data) occur in
the cases of the Gd, Er and Yb isotopes.  Note that the decrease in
the experimental $g$--$\gamma$ transition probabilities is well
consistent with the corresponding increase in the SU(3) splitting.
In this way, the experimental data strongly support the VBM
predictions in the SU(3) contraction limit.

The two-dimensional surface obtained for the theoretical $R_{2}(2)$
ratio, Eq.~(\ref{R2E2}), is shown in Fig.~5.  We see that $R_{2}(2)$
gradually decreases with $\l$ and $\m$ and trends towards the finite
values, obtained in the limiting cases (i) and (ii) [See
Eqs.~(\ref{r2lim1}) and (\ref{r2lim2})]. In Fig.~6 the transition
ratio $R_{2}(L)$ [Eq.~(\ref{R2(L)})] is plotted as a function of the
quantum number $\l$ for $L=2,4\dots 12$. In the case
$\l\rightarrow\infty $, $\m =2$ it gradually goes to the Alaga values
for the corresponding  angular momenta (Fig. 6(a), see also
Eq.~(\ref{limB2L})). In case (ii) $R_{2}(L)$ trends to finite values
which (in the numerically investigated $\l$- range) exhibit a
complicated behavior as a function of $L$ (Fig. 6(b)). In both
limiting cases the lack of dependence on the Hamiltonian parameters
indicates the complete reduction of the SU(3) symmetry. The
experimental data on $R_{2}(2)$, given in Table~1, show a slightly
expressed trend of decreasing towards the midshells, but one could
not draw any definite conclusions about the systematic behavior of
this quantity.

In Fig.~7 the transition ratio $R_{3}(L)$ [Eq.~(\ref{R3(L)})] is
plotted as a function of the quantum number $\l$ for the odd angular
momenta $L=3,5\dots 11$. For $\l\rightarrow\infty $, $\m =2$ it goes
to the corresponding Alaga values in a way similar to the
$R_{2}(L)$- ratio (Fig. 7(a), see also Eq.~(\ref{limB3L})). In the
second direction, (ii), $R_{3}(L)$ saturates to finite
values (Fig. 7(b)). It is clear that towards the SU(3) contraction
limit the B(E2) transition characteristics of the odd $\gamma$ band
states should be consistent with the even angular momentum ones.

In order to assess quantitatively the results presented so far, we
provide a numerical analysis of the SU(3) multiplets in $(\l ,\mu )$-
plane on the basis of the experimental energy and transition ratios
given in Table~1. More precisely we determined the quantum numbers
$\l$ and $\mu$ together with the Hamiltonian parameters $g_{1}$
$g_{2}$ and $g_{3}$, by fitting them in the numerical procedure of
Ref.~\cite{MDRRB97} so as to reproduce the experimental $g$- and
$\gamma$- band levels up to $L=8$, for lantanides and $L=18$, for
actinides together with the values of the experimental ratios
$\Delta E_{2}$, $R_{1}(2)$ and $R_{2}(2)$. (Only for two nuclei,
$^{160}$Gd and $^{162}$Dy, the $R_{1}(2)$ ratios have not been used
in the fits due to the uncertainty of the experimental data.)

The ``favored'' values of the quantum numbers $\l$ and $\mu$ obtained
for the various isotopes are given in the fourth column of Table~1.
Generally the quantum number $\l$ vary in the range $14\leq\l\leq
68$, while $\mu$ obtains the values $2\leq\mu\leq 6$.

So, one finds that while  $\mu$ is closed in narrow limits,
the quantum number $\l$ exhibits a well pronounced systematic
behavior. The favored $\l$-values as well as the SU(3) splitting
ratio $\Delta E_{2}$ increase with the increase of the valence pair
number $N$, i.e. towards the middle of the valence shells in
rotational nuclei. For example, the small splitting observed in the
nuclei $^{152}$Sm ($\Delta E_{2}=7.9$), $^{154}$Gd ($\Delta
E_{2}=7.1$) and $^{162}$Er ($\Delta E_{2}=7.8$) which are situated
near the beginning of the respective group of rotational isotopes, is
associated with the small $\l$- values, $\l =14-16$ and the
relatively large interband transition ratios $R_{1}(2)=0.07-0.08$. On
the other hand, for the middshell nuclei $^{172,174}$Yb with large
splitting values, $\Delta E_{2}=18-20$ and small $R_{1}(2)\sim 0.01$,
we obtain large $\l$-values, $\l\sim 60-70 $. Also, large $\l$
values, $\l\sim 60$ have been obtained for the $^{234,238}$U isotopes
with $\Delta E_{2}=20-22$. Well pronounced systematic behavior of the
quantum number $\l$ is observed in the Er and Yb isotopes.

The above results are consistent with the numerical analyses carried
out in \cite{MDRRB97}. It should be mentioned that the involvement of
the interband transition ratio $R_{1}(2)$ in the present fits leads
to an increase in the quantum number $\m$ above the $\m=2$ values.
Actually, some trend of small increase in $\m$ (up to $\m =6$) with
the increase of $\l$ is indicated for various isotope groups (See, for
example the Yb isotopes in Table~1). Nevertheless, in almost all
nuclei under study the quantum number $\l$ is essentially larger than
$\m$, which is natural for the well deformed nuclei. Several
exceptions are observed in the nuclei far from the midshell region,
such as $^{158}$Dy (with $(\l ,\mu )=(16,6)$) and $^{186}$W (with
$(\l ,\mu )=(24,10)$) where the interband transition ratios are very
large $R_{1}(2)\sim 0.1-0.2$.

The above quantitative considerations show that the changes in the
SU(3) characteristics of nuclei (especially the quantum
number $\l$) towards the middle of given rotational region could be
associated with the corresponding decrease in the $g$--$\gamma$ band
mixing interaction towards the SU(3) contraction limits.  In terms of
our study the strong $g$--$\gamma$ splitting, observed near
the middle of rotational regions, corresponds to the weak
mutual perturbation of the bands. This is consistent with the
respectively good rotational behavior of the $g$-band, which in this
case could belong to a separate SU(3) multiplet.  [See the
experimental energy ratios, $R_{4}=E_4^g/E_2^g$, given in Table~1.]

We remark that present analyses are based mainly on data in the
rare earth region.  Actually Table~1 includes only nuclei for which
the $g$--$\gamma$ interband transition probabilities are measured.
That is why only four actinides (${}^{230,232}$Th, ${}^{234,238}$U)
are considered. Nevertheless, they give an indication for similar
behaviors of the splitting and the interband transition ratios as the
ones in rare earth nuclei.  On the other hand, the generally stronger
energy splitting, observed in the actinide region (See also
\cite{Sakai}), suggests a generally weaker $g$--$\gamma$ coupling
compared to the rare earth nuclei.

We are now able to discuss the physical significance of the
considered SU(3) contraction limits as well as to depict the
physically meaningful directions in the $(\l ,\m )$- plane, which
could be appropriate for studying the transition between the
different band coupling schemes.  The theoretical analyses and
experimental data show that the limiting case (i)
($\l\rightarrow\infty$, with $\m$ finite) has a rather clear physical
interpretation.  It is consistent with the observed continuous
increase of the $g$--$\gamma$ band splitting and the corresponding
continuous decrease of the interband transition probabilities towards
the midshell regions in rare earth nuclei.  The limiting case (ii)
($\l = \m \rightarrow\infty $) does not have any similar direct
interpretation.  It suggests finite values for the splitting and the
interband transition probabilities, while the bands do not interact
in the framework of SU(3) symmetry.  In addition, it is well known
that the case $\l = \m$ does not correspond to deformed nuclei, for
which the inequality $\m < \l$ is satisfied.  Nevertheless the study
of this limit is useful from the following viewpoint:  It implies
that the strong suppression of the band interaction as well as the
transition between the different band coupling schemes could be
realized at reasonable (finite) SU(3) splitting.  Based on the above
considerations, we deduce that the possibly interesting physically
meaningful directions in the $(\l ,\m )$--plane should be associated
with a consistent increase in the quantum numbers $\l$ and $\m$.
Thus, any particular direction of interest could be easily estimated
by using its intermediate behavior between the two considered
limiting cases.

Some discussion concerning the Interacting Boson Model (IBM)
classification scheme \cite{IA} is appropriate at this point.  In
Ref. \cite{MDRRB97} it has been suggested that for deformed nuclei
both the VBM scheme (with $g$--$\gamma$ band coupling) and the IBM
one (with $\beta$--$\gamma$ band coupling) could be considered as
complementary schemes.  It has also been pointed out that the SU(3)
scheme of the VBM is naturally applicable to nuclei with weak energy
splitting, while strong splitting invokes the SU(3) scheme of the
IBM, in which the $g$- band is situated in a separate irrep.
Furthermore, the theoretical results and the experimental data given
in the present work suggest that the VBM band coupling scheme is more
appropriate near the ends of the rotational regions, while in the
midshell regions the coupling scheme of the IBM is realized.  In this
respect the detailed comparison of both band-coupling mechanisms
would be of interest.  For example, the analytic expressions for the
$g$--$\gamma$ interband transition probabilities, obtained in the
framework of the IBM in \cite{PvI83}, would be useful.  [See Eqs.~(5)
and (6) of \cite{PvI83}.] They give a behavior of the transition
ratios $R_{1}(L)$, $R_{2}(L)$ and $R_{3}(L)$ [Eqs.~(\ref{R1(L)}),
(\ref{R2(L)}) and (\ref{R3(L)})] in the infinite valence pair number
limit ($N\rightarrow\infty$) similar to the behavior obtained in the
limiting case (i) ($\l\rightarrow\infty $, with $\m =2$) [Eqs.
(\ref{limB1L}), (\ref{limB2L}) (\ref{limB3L})] of the present VBM
scheme.

As an extension of the present studies it would be worthwhile to
examine, in a similar way, the link between the $\gamma$- and the
$\beta$- band. Furthermore, besides the VBM scheme, one could refer
in this case to the modifications of the IBM in which higher-order
terms conserving the SU(3) symmetry are added \cite{BMI85}.  The
consistent study (within both models) of the ways in which the SU(3)
symmetry is reduced could give important information about the
rearrangement of rotational bands into different SU(3) irreps.

In the above context, we emphasize that the analyses implemented in
the presented paper give a general prescription to handle the fine
behavior of the band coupling interactions in any collective
algebraic scheme in heavy deformed nuclei. Actually, the group
contraction process should play a major role in a  transition between
two different band coupling schemes. The transition from the compact
SU(3) group to the non-compact T$_{5}\wedge$SO(3) rotor group could
be considered as a starting point in a process of reconstruction of
various multiplets in a more general symplectic group of dynamical
symmetry. (It is interesting to mention that  the meaning of SU(3)
contraction has been also discussed (though in a rather different
aspect) in reference to a possible phase transition between a
superconductor and rigid rotor collective motion of nuclei
\cite{BRW98}.)

\section{Conclusions}

We have derived analytic expressions for the energies and $B(E2)$
transition probabilities in the ground- and $\gamma$-band states of
even deformed nuclei within the Vector-Boson Model with SU(3)
dynamical symmetry. On this basis we applied both analytic and
numeric analyses to examine the behavior of the corresponding energy
splitting and B(E2) transition ratios in the two SU(3) contraction
limits of the model,
    (i) ($\l\rightarrow\infty $, with $\m$ finite),  and
    (ii) ($\l =\m \rightarrow\infty $).
It has been shown that in both limits the $g$--$\gamma$ band mixing
decreases asymptotically to zero.  In case (i) this is associated
with the corresponding continuous increase in the splitting of the
multiplet and the rapidly vanishing $g$--$\gamma$ interband
transition link.  Case (ii) gives finite values for the energy
splitting and the interband transition ratios which, however, should
not be associated with any band coupling.  The latter result implies
that a strong reduction of the band interaction could be possible at
finite SU(3) splitting.  Thus, the present analyses outline the
possible directions in the $(\l ,\m )$-plane in which the
$g$--$\gamma$ band coupling is reduced.

The experimental data on the ground- and $\gamma$- band states in
deformed even--even nuclei show clearly a pronounced increase in the
$g$--$\gamma$ band splitting and a corresponding decrease in the
interband transition probabilities towards the midshell regions.
They suggest that the SU(3) contraction effects in the $g$--$\gamma$
band coupling scheme should be sought in the best rotational nuclei,
in which the mutual perturbation of the bands is weak. So,
the experimental data and their quantitative estimation in the
VBM framework strongly support our theoretical analyses.

Based on the presented investigation, we conclude that the transition
from the $g$--$\gamma$ band coupling scheme to a scheme in which the
$g$-band is situated in a separate irrep should be realized towards
the midshell regions.  In this respect the complementarity of the
classification schemes of the Vector-Boson Model with SU(3) dynamical
symmetry and the IBM becomes clear.  The consistent study of the
rearrangement of collective bands in deformed nuclei, including the
$\beta$- excited bands, will be the subject of forthcoming work.

\bigskip
{\Large{\bf Acknowledgments}}
\medskip

The authors are thankful to P. Van Isacker for stimulating
discussions and D. N. Kadrev for the help in collecting the
experimental data.
This work has been supported by the Bulgarian
National Fund for Scientific Research under contract no
MU--F--02/98.

\ \ \ \

\begin{table}
\caption{Experimental values of the energy splitting
(Eq.~(\protect\ref{eq:split}), column 5) and B(E2)-interband
transition ratios in the $2_g$ and $2_\gamma$ states
(Eq.~(\protect\ref{R1E2}) in column 6, Eq.~(\protect\ref{R2E2}) in
column 7) of deformed rare earth nuclei and actinides.
The corresponding favored SU(3)- quantum numbers $(\l ,\mu )$ are
given in column 4. The valence pair number, $N$, as well as the
$R_{4}=E_4^g/E_2^g$ energy ratio are also given in columns 2 and 3
respectively. Data are taken from \protect\cite{Sakai}, for the
energies and from the Refs. in the last column, for the transition
probabilities.}

{\small
\begin{center}
\begin{tabular}{cccccccc}
\hline
Nucl. & N & $R_4=\frac{E_4^g}{E_2^g}$ & $(\l ,\mu )$ &
$\Delta E_{2}=\frac{E_{2}^{\gamma}-E_{2}^{g}}{E_{2}^{g}}$ &
$R_{1}(2)=\frac{2_{\gamma}\rightarrow
2_g}{2_g\rightarrow 0_g}$&
$R_{2}(2)= \frac{2_{\gamma}\rightarrow 2_g}
{2_{\gamma}\rightarrow 0_g}$ & Ref. \\
\hline
${}^{152}\rm Sm$& 10& 3.009& (14,4)&  7.915&  0.065 (5) &
2.56 (26)& \cite{NDS152} \\
${}^{154}\rm Sm$& 11& 3.253& (58,6)& 16.565&  0.022   &
1.35& \cite{NDS154}, \cite{NDS182} \\
${}^{154}\rm Gd$& 11& 3.015& (14,4)& 7.093&  0.083 (6) &
2.15 (24)& \cite{NDS154} \\
${}^{156}\rm Gd$& 12& 3.239& (24,4)& 11.969&  0.039 (3) &
1.56 (12)& \cite{NDS156} \\
${}^{158}\rm Gd$& 13& 3.288& (28,4)& 13.932&  0.029 (4) &
1.71 (41)& \cite{NDS158} \\
${}^{160}\rm Gd$& 14& 3.298& (22,2)& 12.128&  $\leq 10^{-3}$ &
1.69 (19)& \cite{NDS160}, \cite{Raman87} \\
${}^{158}\rm Dy$& 13 & 3.207& (16,6)& 8.567&  0.103 (23) &
3.22 (133)& \cite{NDS158} \\
${}^{160}\rm Dy$& 14 & 3.270& (16,2)&10.131&  0.028 (9) &
1.93 (112)& \cite{NDS160} \\
${}^{162}\rm Dy$& 15 & 3.294& (16,2)&10.007& $\leq 10^{-3}$ &
1.66& \cite{NDS162} \\
${}^{164}\rm Dy$& 16 & 3.301& (16,2)&9.379&  0.038 (4) &
2.00 (38)& \cite{NDS164} \\
${}^{162}\rm Er$& 13 & 3.229& (16,4)&7.822&  0.067 (11) &
2.37 (35)& \cite{NDS162} \\
${}^{164}\rm Er$& 14 & 3.277& (18,4)&8.412&  0.052 (7) &
2.19 (48)& \cite{NDS164} \\
${}^{166}\rm Er$& 15 & 3.289& (20,4)&8.751&  0.045 (5) &
1.76 (26)& \cite{NDS166} \\
${}^{168}\rm Er$& 16 & 3.309& (22,4)&9.291&  0.0410 (3) &
1.80 (12)& \cite{NDS168} \\
${}^{170}\rm Er$& 17 & 3.310& (30,4)&10.858&  0.034 (7) &
1.93 (36)& \cite{NDS170} \\
${}^{168}\rm Yb$& 14 & 3.266& (14,2)&10.218&  0.046 (6) &
2.09 (72)& \cite{NDS168} \\
${}^{170}\rm Yb$& 15 & 3.293& (18,2)&12.594&  0.024 (6) &
1.78 (77)& \cite{NDS170} \\
${}^{172}\rm Yb$& 16 & 3.305& (58,4)&17.602& 0.011 (3)&
1.45 (65)& \cite{NDS172}, \cite{NDS182} \\
${}^{174}\rm Yb$& 17 & 3.310& (68,6)&20.356& 0.012 (3) &
2.40 (94)& \cite{NDS174} \\
${}^{176}\rm Yb$& 16 & 3.308& (42,4)&14.358& 0.018 (4) &
1.94 (70)& \cite{NDS176} \\
${}^{174}\rm Hf$& 15 & 3.268& (36,6)&12.481& 0.049 (12)&
1.54 (63)& \cite{NDS174} \\
${}^{178}\rm Hf$& 15 & 3.291& (30,4)&11.604&  0.028 (2) &
1.18 (19)& \cite{NDS178} \\
${}^{182}\rm W $& 13 &3.291& (18,4)&11.203&   0.053 (6) &
1.90 (19)& \cite{NDS182} \\
${}^{184}\rm W $& 12 &3.274& (16,4)&7.123&   0.071 (5)   &
1.91 (19)& \cite{NDS184} \\
${}^{186}\rm W $& 11 &3.242& (24,10)& 5.030&   0.181 (13) &
2.27 (32)& \cite{NDS186} \\
${}^{230}\rm Th $& 11 &3.272& (24,4)&13.586&  0.028 (5)  &
1.83 (53)& \cite{NDS230} \\
${}^{232}\rm Th $& 12 &3.284& (26,4)&14.893&   0.036 (6)  &
2.73 (62)& \cite{NDS232} \\
${}^{234}\rm U $& 13 &3.296& (58,6)&20.304&   0.021 (5)   &
1.69 (69)& \cite{NDS234} \\
${}^{238}\rm U $& 15 &3.304& (60,6)&22.614&   0.019 (1)   &
1.75 (17)& \cite{NDS238} \\
\hline
\end{tabular}
\end{center}
}
\end{table}

\newpage
    \begin{center}
    {\bf Figure Captions}
    \end{center}
    \bigskip\bigskip

\noindent
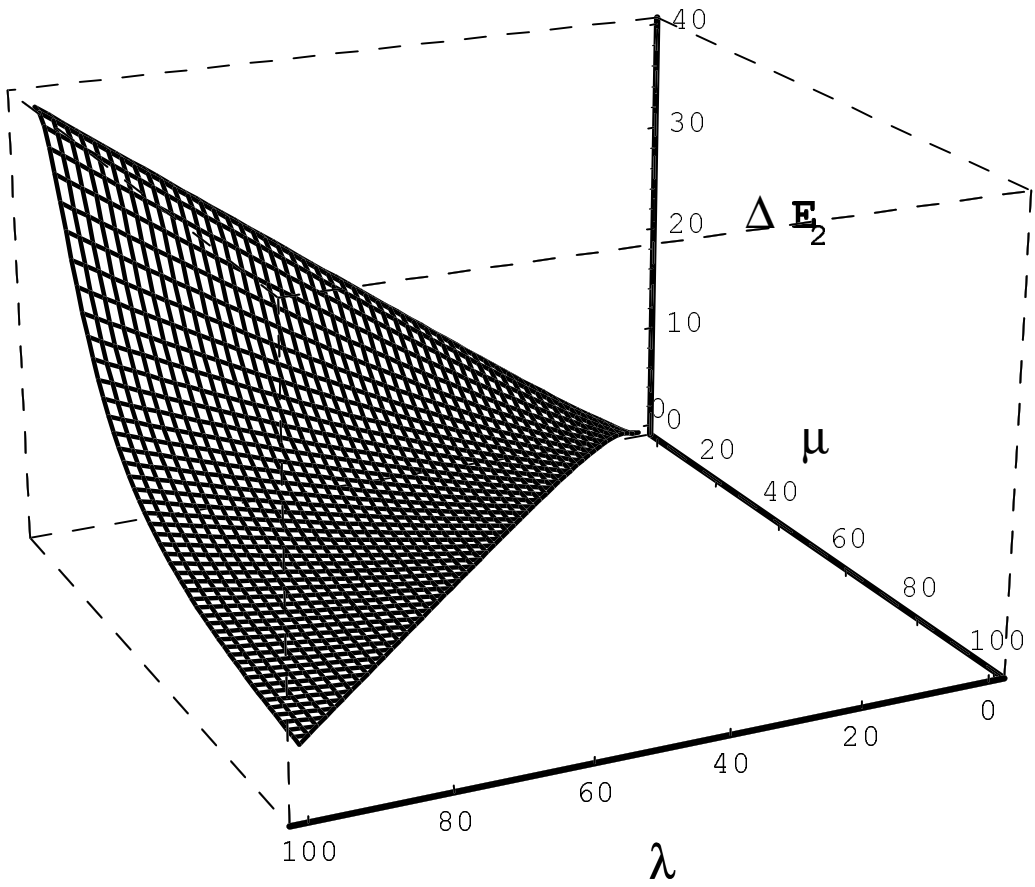
{\bf Figure 1.} The theoretical energy splitting ratio  $\Delta
E_{2}$ [Eq.~(\ref{anspl})] is plotted as a two-dimensional function
of the quantum numbers $\l$ and $\m$ for $g_{1}=1$, $g_{2}=-0.2$ and
$g_{3}=-0.25$.
\medskip

\noindent
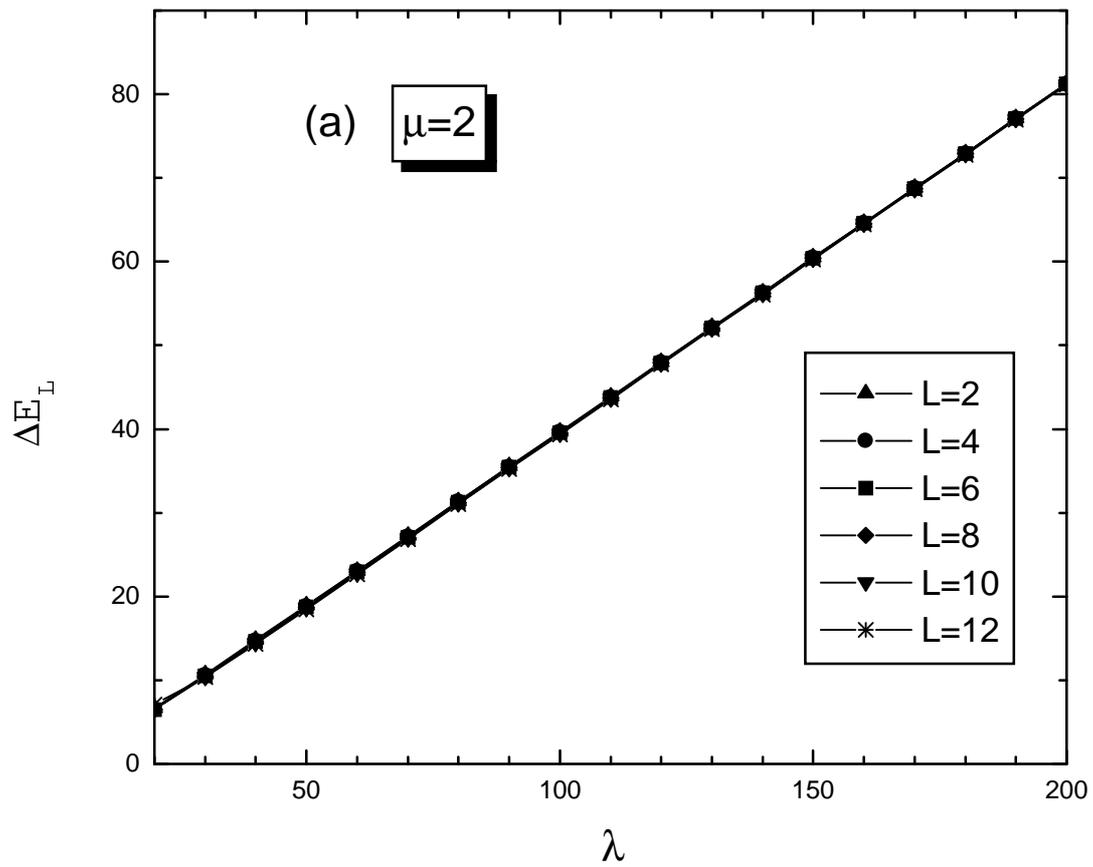
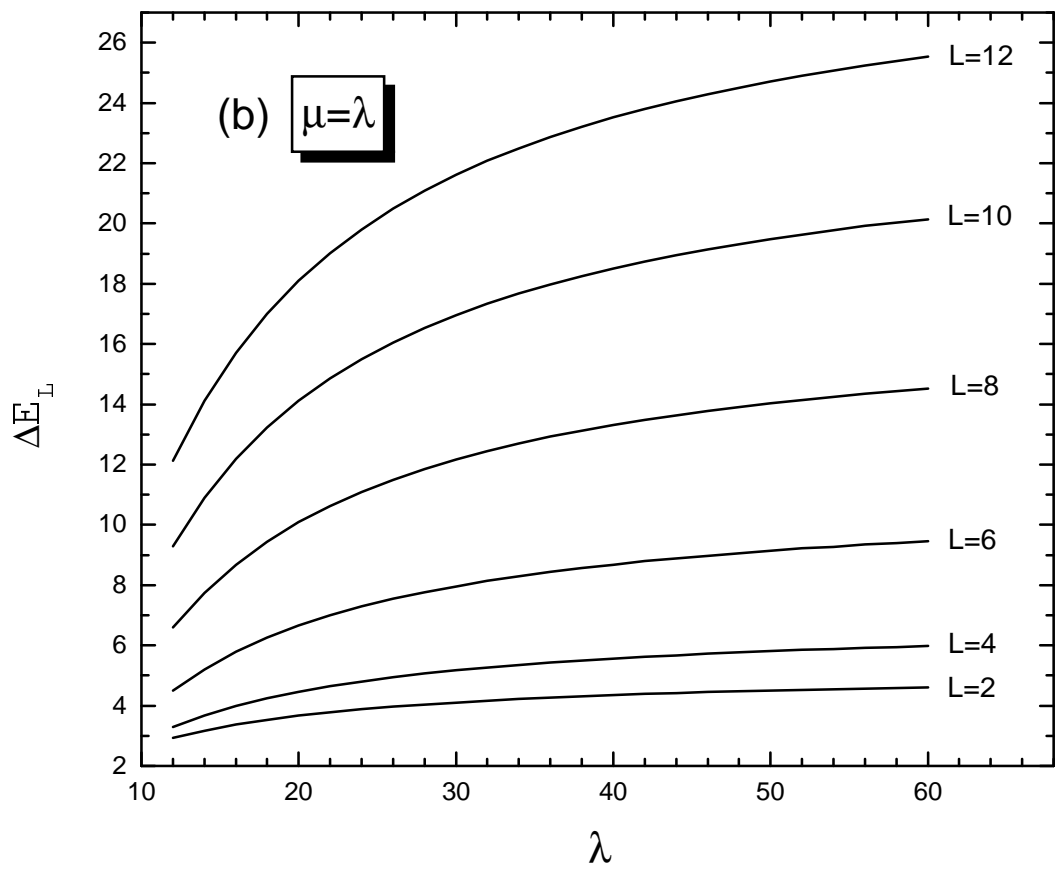
{\bf Figure 2.} The theoretical energy splitting ratio  $\Delta
E_{L}$ [Eq.~(\ref{eq:splitL})] is plotted as a function
of the quantum number $\l$ for $L=2,4\dots 12$ with  $g_{1}=1$,
$g_{2}=-0.2$ and $g_{3}=-0.25$ in the cases: (a) $\m=2$;
(b) $\m =\l$.
\medskip

\noindent
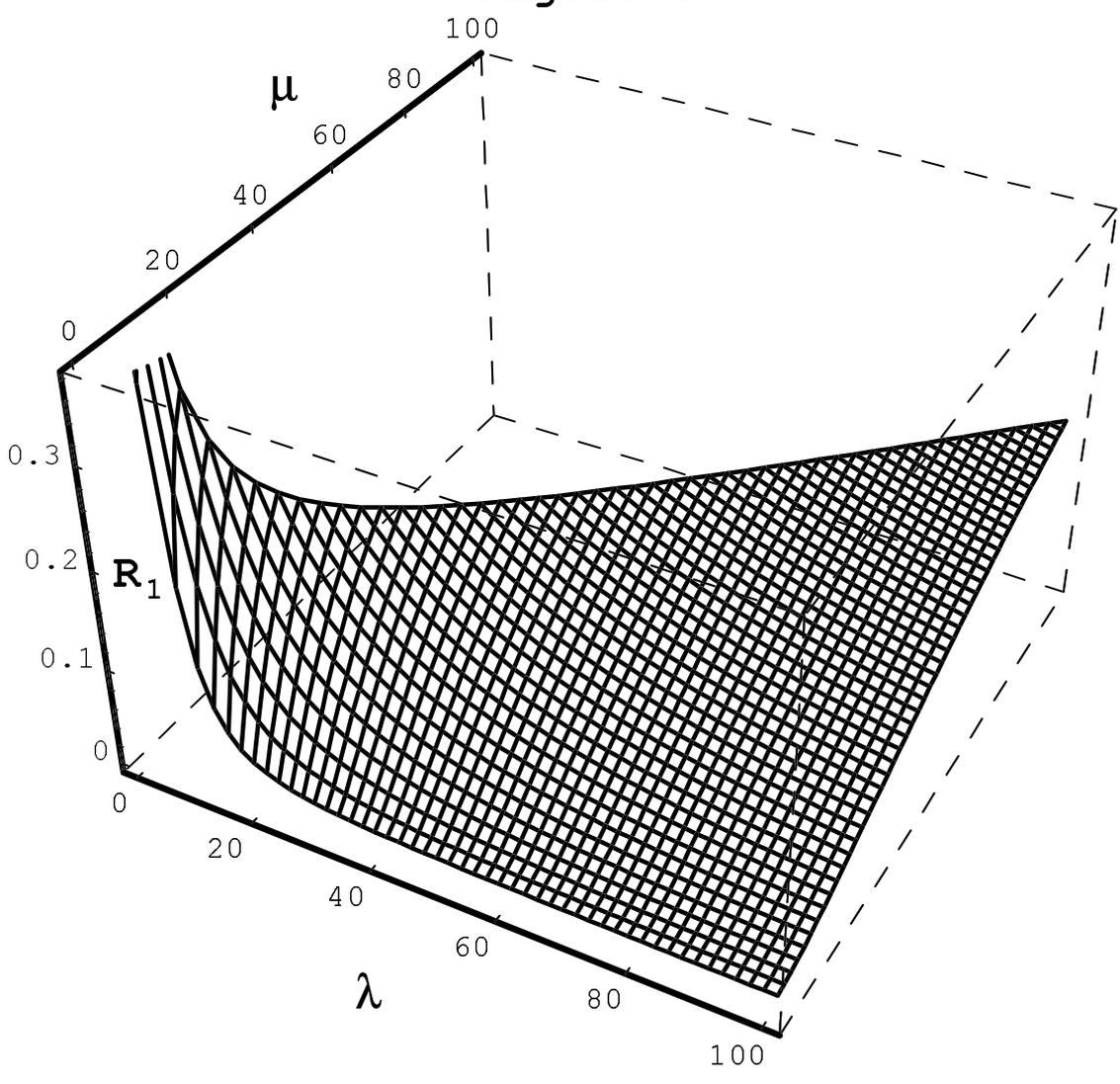
{\bf Figure 3.} The theoretical $R_{1}(2)$ ratio [Eq.~(\ref{R1E2})] is
plotted as a two-dimensional function of the quantum numbers $\l$ and
$\m$ for $g_{2}=-0.2$ and $g_{3}=-0.25$.
\medskip\medskip

\noindent
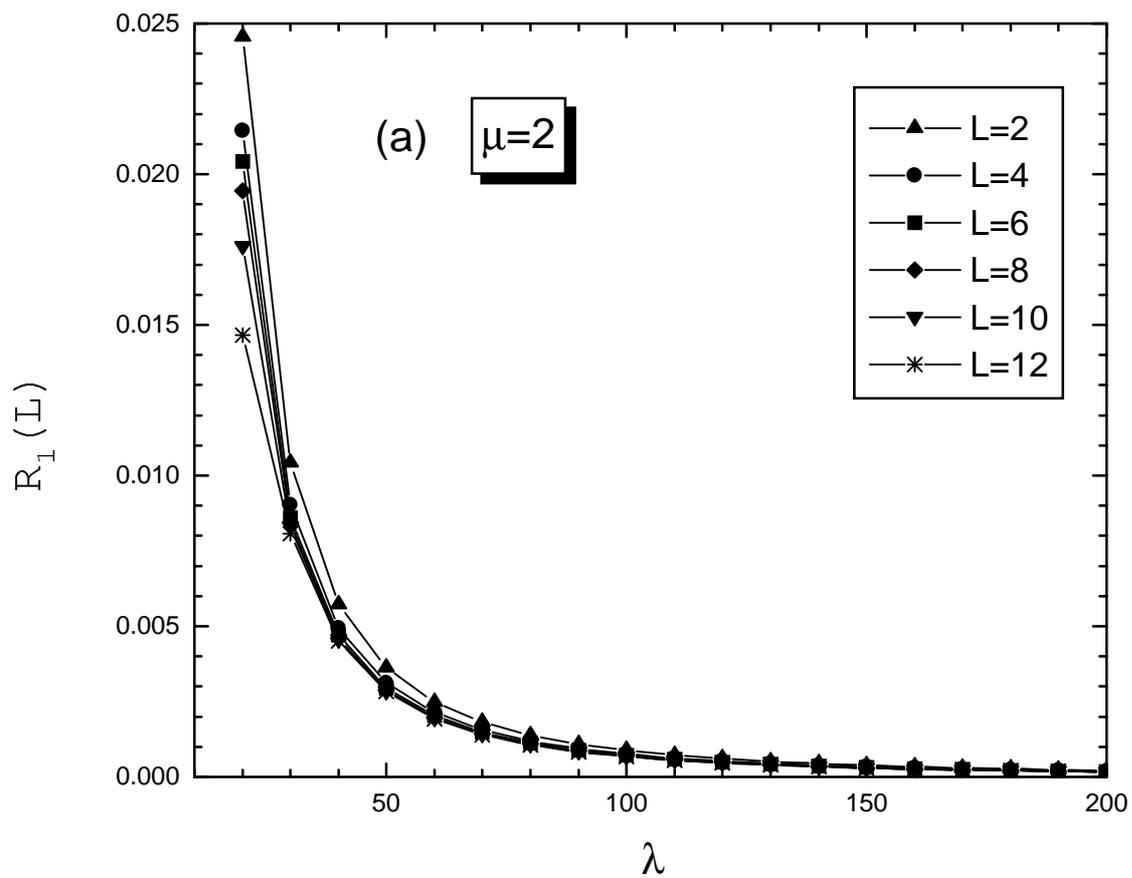
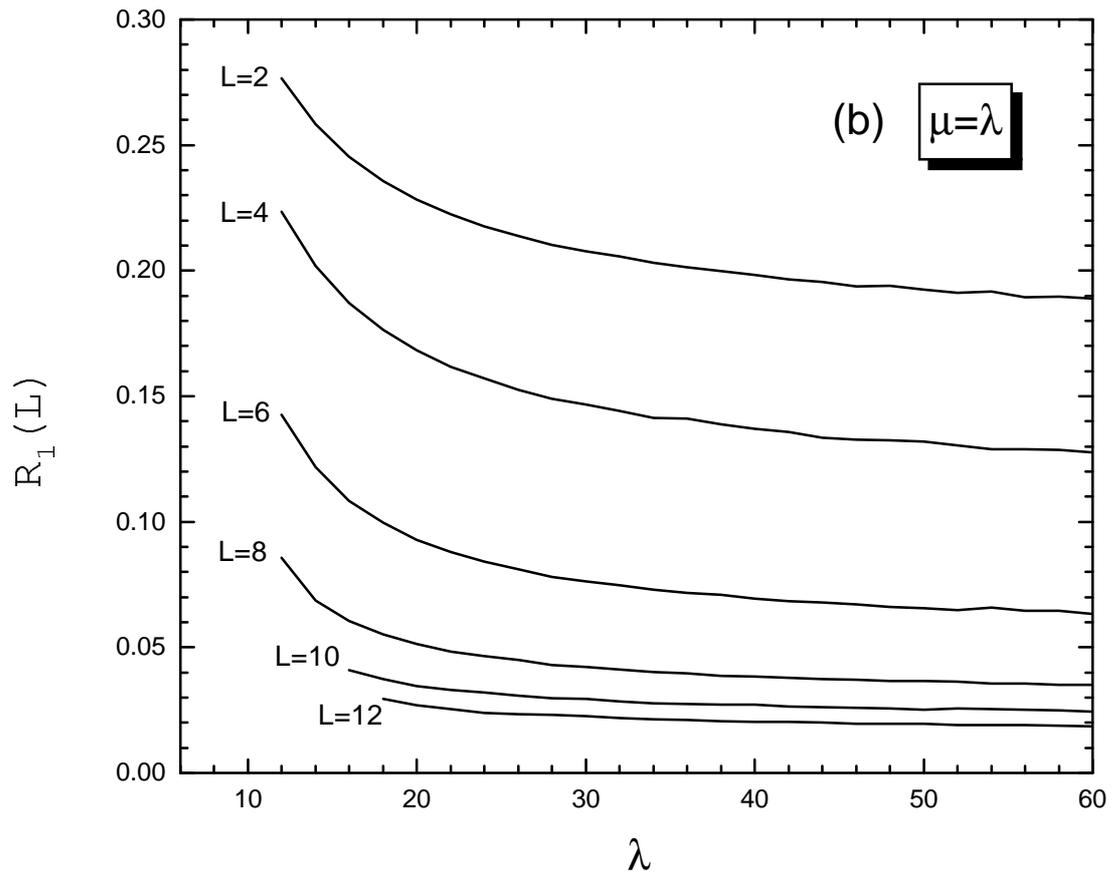
{\bf Figure 4.} The theoretical $R_{1}(L)$ ratio [Eq.~(\ref{R1(L)})]
is plotted as a function of the quantum number $\l$ for $L=2,4\dots
12$ with  $g_{2}=-0.2$ and $g_{3}=-0.25$ in the cases: (a) $\m=2$;
(b) $\m =\l$.
\medskip

\noindent
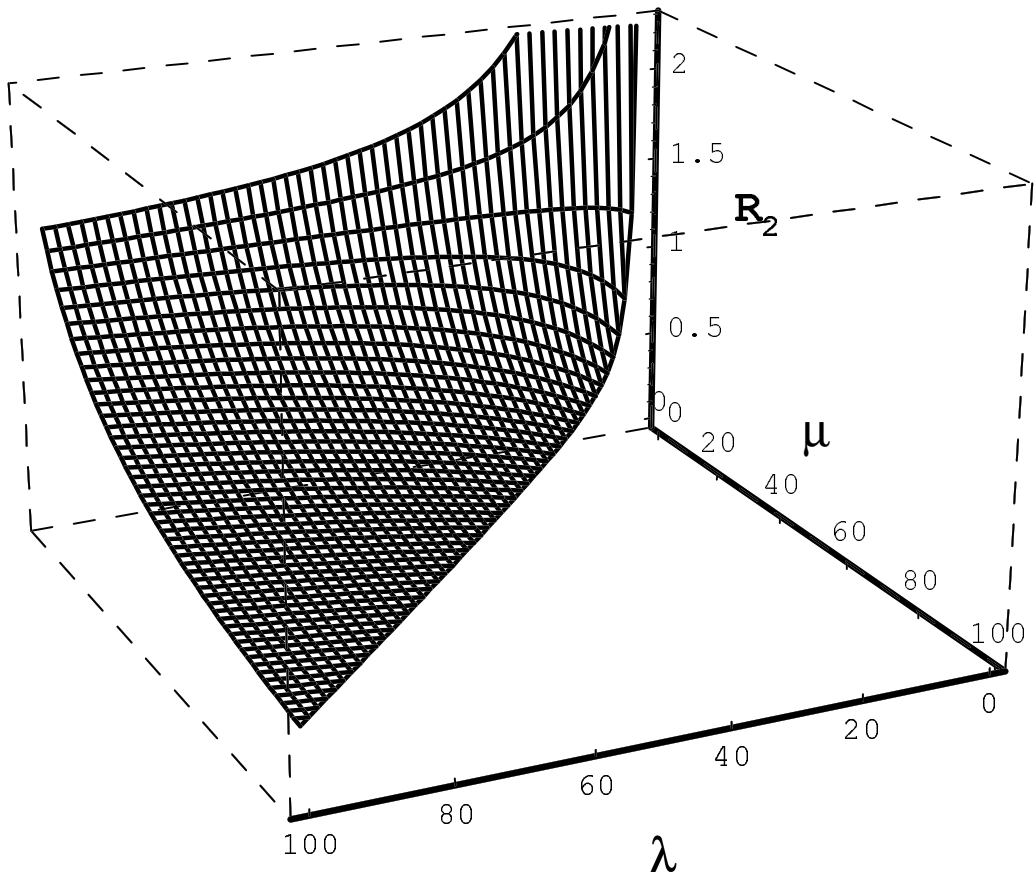
{\bf Figure 5.} The same as Fig. 3 but for the theoretical
$R_{2}(2)$ ratio [Eq.~(\ref{R2E2})].
\medskip

\noindent
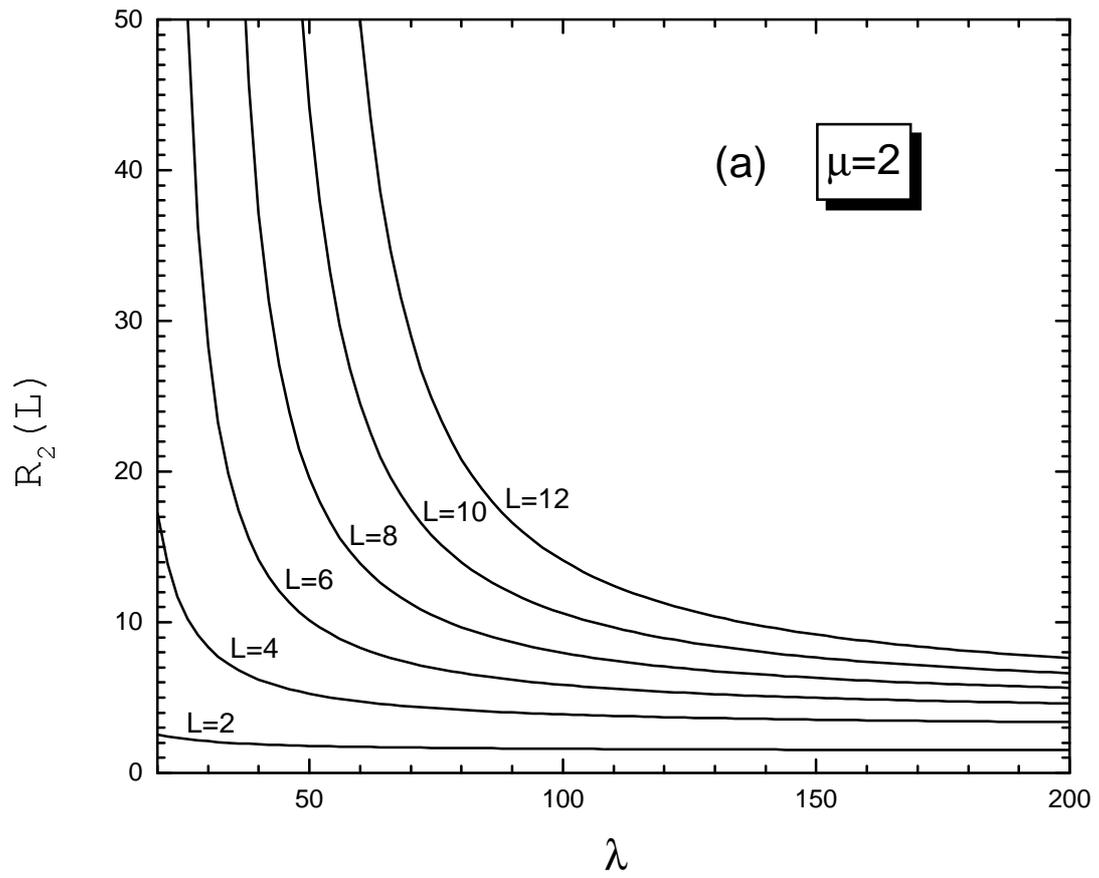
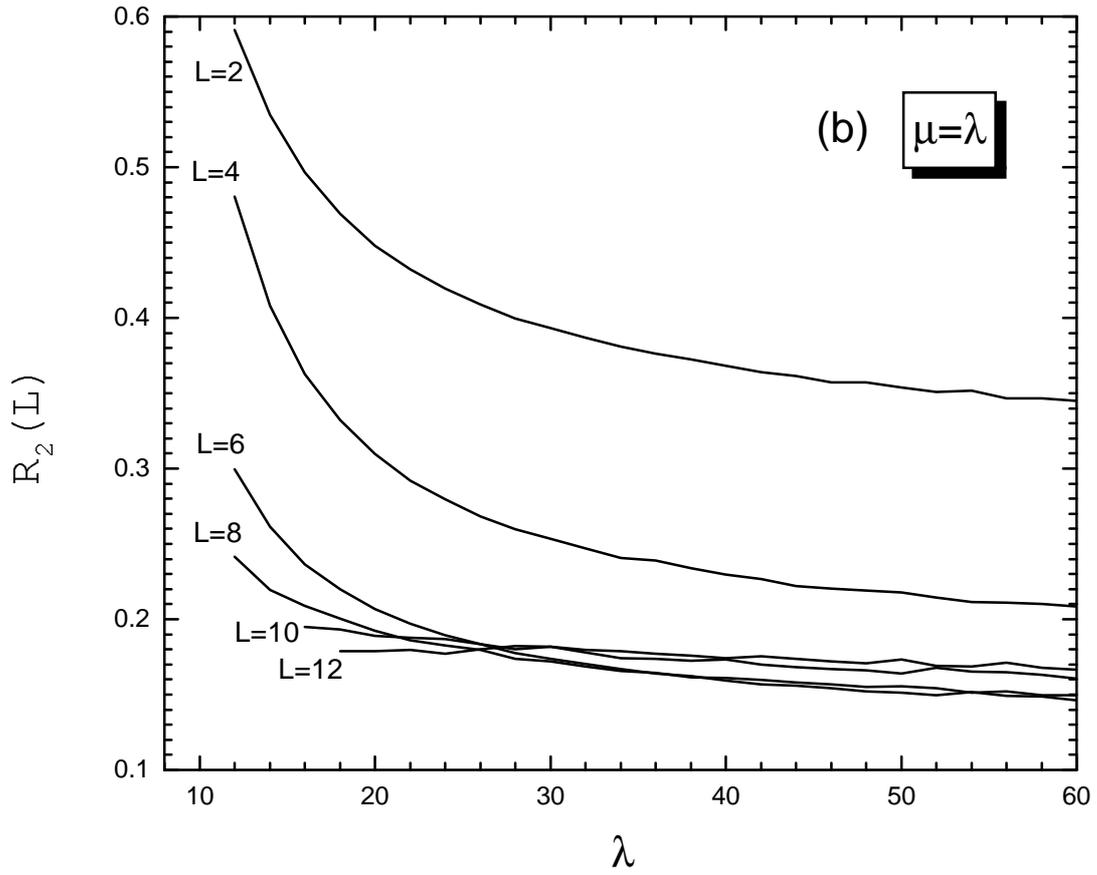
{\bf Figure 6.} The same as Fig. 4 but for the theoretical
$R_{2}(L)$ ratio [Eq.~(\ref{R2(L)})].
\medskip

\noindent
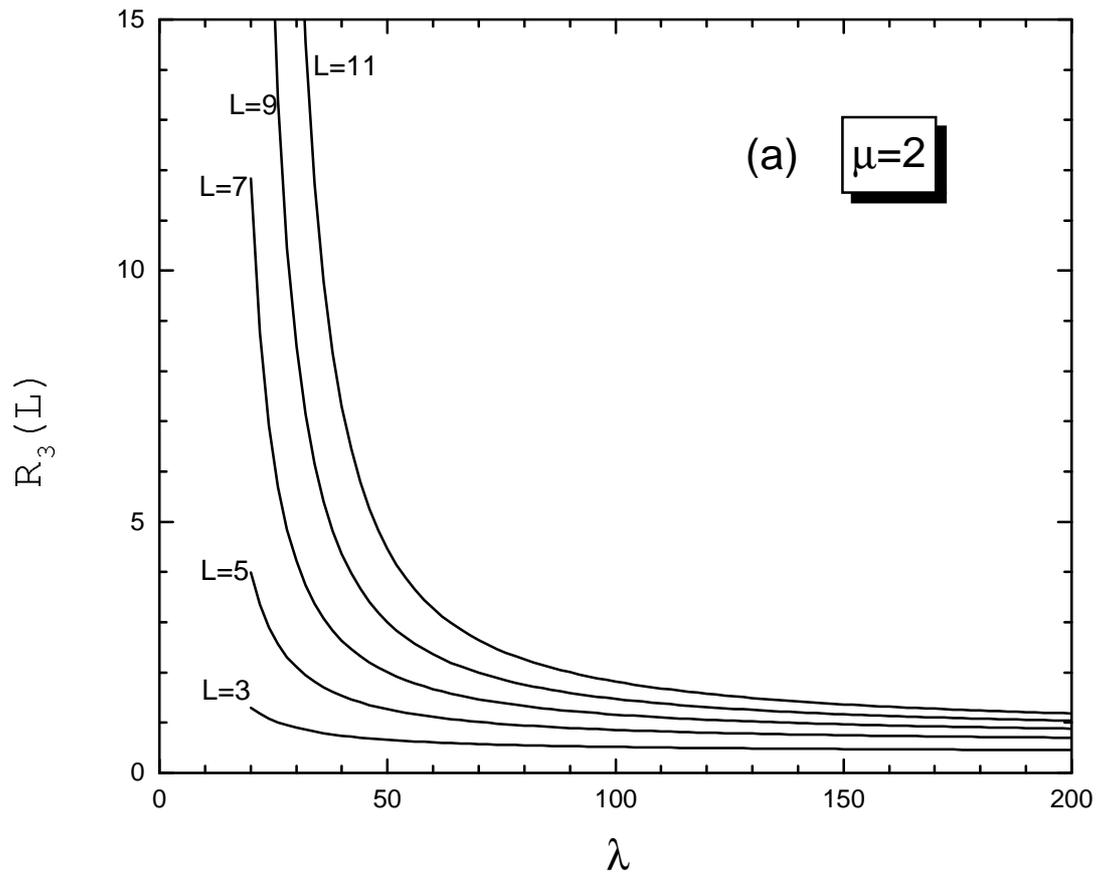
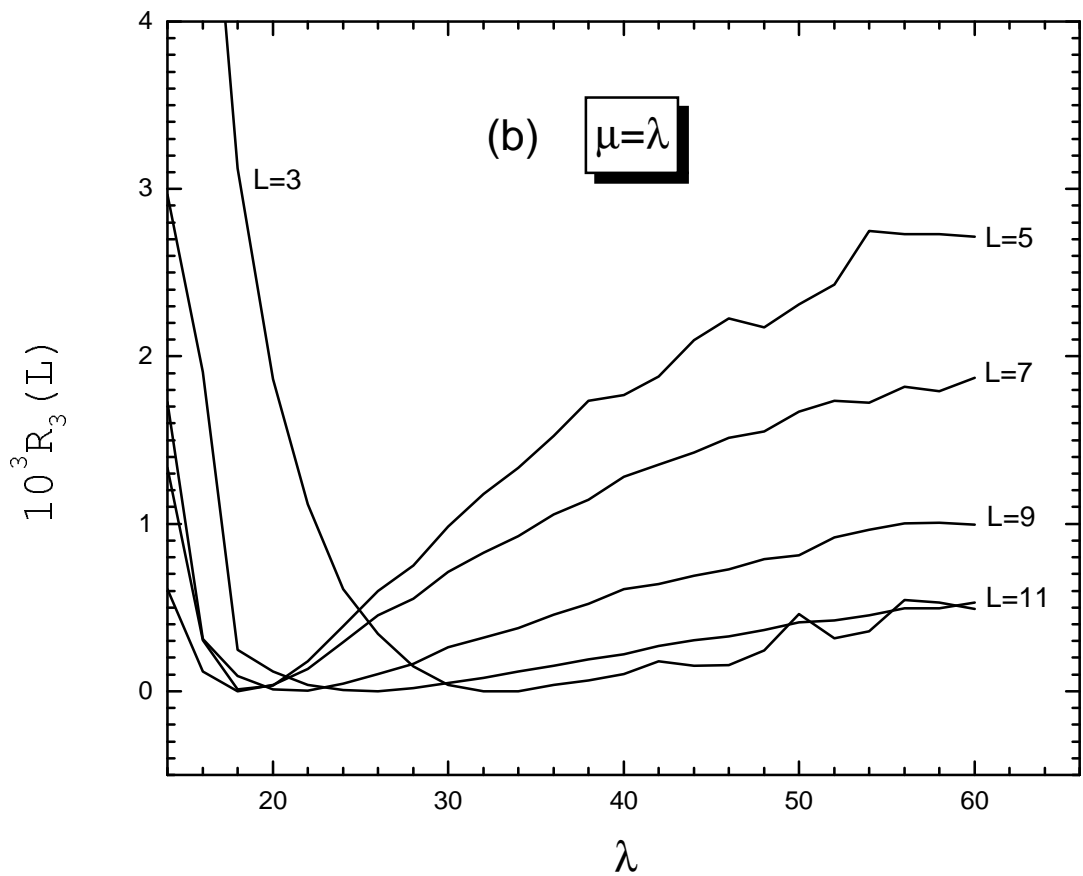
{\bf Figure 7.} The theoretical $R_{3}(L)$ ratio [Eq.~(\ref{R3(L)})]
is plotted as a function of the quantum number $\l$ for $L=3,5\dots
11$ with $g_{2}=-0.2$ and $g_{3}=-0.25$ in the cases: (a)
$\m=2$; (b) $\m =\l$.

\end{document}